\documentclass[useAMS,usenatbib,usegraphicx]{mn2e}
\usepackage{amsmath,amsfonts,amssymb,url}
\usepackage{pslatex}

\usepackage{txfonts}

\defcitealias{nautical2013astronomical}{The Astronomical Almanac}

\def\idm#1{{\mbox{\scriptsize #1}}}

\newcount\exno

\usepackage{color}
\definecolor{myred}{rgb}{0.55,0.05,0.05}
\definecolor{myblue}{rgb}{0.1,0.0,0.5}  
\definecolor{mybrown}{rgb}{0.9,0.4,0.3} 
\newcommand\ed[1]{{\global\advance\exno by 1\
\color{myblue}$\spadesuit$({\bfseries\the\exno}).
\color{myblue}\bfseries\em #1}}

\newcommand\hide[1]{}
\newcommand\Chi{{\chi^2_\nu}}
\def\tv#1{{\pmb #1}}
\def\tvlm{TVLM\,513}

\title[Astrometry and astrophysics of \tvlm{} revisited]
{Physical properties and astrometry of radio-emitting brown dwarf
TVLM\,513-46546 revisited}
\author[M. Gawro\'nski, K.~Go\'zdziewski \& K. Katarzy\'nski]%
{Marcin P. Gawro\'nski, Krzysztof Go\'zdziewski \& Krzysztof Katarzy\'nski\\
Centre for Astronomy, Faculty of Physics, Astronomy and Informatics,
Nicolaus Copernicus University, Grudziadzka 5, 87-100 Toru\'n, Poland}
\begin{document}

\date{Accepted .... Received .....; in original form ....}

\pagerange{\pageref{firstpage}--\pageref{lastpage}} \pubyear{2014}

\maketitle

\label{firstpage}

\begin{abstract}
We present multi-epoch astrometric observations of the M9 ultra-cool dwarf
TVLM\,513-46546 that is placed at the brown dwarf boundary. The new observations
have been performed with the European VLBI Network (EVN) at 6~cm band. The
target has been detected at 7~epochs spanning three years, with measured
quiescent emission flux in the range $180-300\,\mu$Jy. We identified four
short-duration flaring events ($0.5-2\,$mJy) with very high circular
polarization ($\sim75\%$~--\,$100\%$). Properties of the observed radio flares
support the physical model of the source that is characterized by the electron
cyclotron maser instability responsible for outbursts of radio emission.
Combined with Very Long Baseline Array (VLBA) earlier data, our detections make
it possible to refine the absolute parallax $\pi=93.27^{+0.18}_{-0.17}$\,mas.
Our measurements rule out TVLM\,513-46546 companions more massive than Jupiter
in orbits with periods longer than $\sim 1$~yr.  
\end{abstract}


%
\section{Introduction}
%
Investigating properties of astrophysical objects in the Solar neighbourhood is
one of the main targets of modern astrophysics regarding stellar and planetary
systems statistics. Since low-mass stellar and sub-stellar population is
dominant in the local volume, much effort is recently devoted to studies of
M-dwarfs and brown dwarfs. These objects are favourable targets to detect their
low-mass companions. However, young M-dwarfs are magnetically active, making it
very difficult to measure their radial velocities (RV) with the precision
required by contemporary planetary surveys ($\sim5-50$ ms$^{-1}$), due to
variable emission lines and broad molecular spectral features. Therefore ongoing
RV surveys focus on M-dwarf samples that are biased towards chromospherically
quiet and old objects
\cite[e.g.,][]{Affer2016,Astudillo2015,Bailey2009,Rivera2005}. Fortunately, in
general M-dwarfs spectral activity has much less impact on astrometric
measurements. 

Astrometric techniques make it possible to reach targets which also could not be
observed by transits. Indeed, the optical/infrared astrometry and direct imaging
has recently revealed two sub-stellar companions orbiting a very low-mass star
and a brown dwarf \cite[e.g.,][]{Sahlmann2016}, see also
\citep{2013A&A...556A.133S,2015A&A...577A..15S,Bowler2015}.

The Very Large Baseline Interferometry (VLBI) technique was already successful
for observations of active M-dwarfs at the radio domain
\citep[e.g.][]{2000A&A...353..569P}. The current performance of the global VLBI
systems make it possible to measure the relative positions with sub-mas
precision even of very weak radio sources ($\simeq 100\,\mu$Jy) and brightness
temperatures in the range of $10^6-10^7$\,K. It provides a unique opportunity to
perform astrometric studies of magnetically active low-mass stars placed in the
Solar neighbourhood. The Radio Interferometric Planet Search (RIPL) conducted
with the Very Long Baseline Array\footnote{\url{http://www.vlba.nrao.edu/}}
(VLBA) demonstrates excellent new VLBI capabilities
\citep{2009ApJ...701.1922B,2011ApJ...740...32B}. 

In this work, we present new results derived in the framework of
Radio-Interferometric Survey of Active Red Dwarfs (RISARD) project
\citep{2013arXiv1309.4639G}. Similar to RIPL, RISARD is an astrometric survey
conducted with the EVN, which is dedicated for observations of very young,
low-mass magnetically active M-dwarfs. Our targets are placed within 10--15\,pc
from the Sun. Here, we focus on a nearby brown dwarf TVLM\,513-46546
\citep[][hereafter \tvlm{}]{1995AJ....110.3014T} placed at the distance of
$10.76\pm0.03$\,pc \citep{Forbrich2013}. 

The paper is structured as follows. After this introduction, we present a
characterisation of the target and its radio-emission in Section~2. Section~3
is devoted to our follow-up EVN astrometric observations of this object. They
extend the time-window by three times to $\sim 7$ years between March, 2008 and
March, 2015 and span 14~epochs, hence they double the number of high-precision
astrometric positions in \citet{Forbrich2013}. A formulation of our improved
astrometric model and the results for all available astrometric measurements are
presented in Sect.~4. We constrain the mass range and orbital period of a
putative sub-stellar or planetary companion of \tvlm{}. The new observations are
useful for the astrophysical characterization of the target discussed in
Sect.~5. The paper ends with conclusions.

%
\section{The \tvlm{} target and its properties}
%

Ultra-cool dwarfs \citep[spectral class M7 and cooler,][]{1997AJ....113.1421K}
attract a great interest as boundary objects between stars and brown dwarfs.
Since the discovery of intense, non--thermal radio emission from stars at the
low-mass end of the main sequence \citep{2001Natur.410..338B,
2002ApJ...572..503B}, the ongoing radio surveys of ultra-cool dwarfs revealed that
$\sim$10\% of these objects are radio luminous
\citep[e.g.][]{2008A&A...487..317A, 2013A&A...549A.131A}. Probably the most
unexpected aspect of the observed radio activity from ultra-cool dwarfs was the
detection of periodic 100\% circularly polarized pulses
\citep{2007ApJ...663L..25H,2008ApJ...684..644H}. Observations by
\citet{2007ApJ...663L..25H} of \tvlm{} showed that electron cyclotron maser
emission is responsible for these 100\% circularly polarized periodic pulses
what implies $\sim$kG magnetic field strengths in a large-scale stable magnetic
configuration. This agrees with the measured $\sim$kG magnetic field
strengths for ultra-cool dwarfs via Zeeman broadening observations
\citep{2007ApJ...656.1121R}. Very recently, \cite{2015Natur.523..568H} detected
radio and optical auroral emissions powered by magnetospheric currents from
ultra-cool dwarf LSR\,J1835+3259 what supports the hypothesis of large-scale
magnetic fields present in ultra-cool dwarfs. Yet it is still unclear which physical
mechanism (incoherent or coherent) is responsible for the quiescent component of
the radio emission. 
The incoherent gyrosynchrotron emission was proposed as the explanation of this
emission by a few authors
\citep[eg.][]{2006ApJ...648..629B,2006ApJ...647.1349O}. Recent detections of
high frequency radio emission from ultra-cool dwarf DENIS\,1048-3956 at 18\,GHz
\citep{2011ApJ...735L...2R} and \tvlm{} at 95\,GHz \citep{2015ApJ...815...64W}
strongly support this explanation at least in these two cases.

The observed radio luminosity of detected ultra-cool dwarfs shows an excess when
compared with the well-known empirical G\"udel--Benz relation between radio and
X-ray luminosity, $L_\idm{radio}$ and $L_\idm{X}$, respectively, which reads as
$L_\idm{radio}/L_\idm{X}\sim 5$ for magnetically active stars
\citep{1993ApJ...405L..63G}. The theoretical model explaining the G\"udel--Benz
relation assumes chromospheric evaporation \citep{2006ApJ...644..484A}. In this
scenario, the X-ray emission results from the heating and evaporation of
chromospheric plasma caused by non-thermal beamed electrons, which produce
gyrosynchrotron radio emission \citep{1968ApJ...153L..59N}. All ultra-cool
dwarfs detected in the radio bands contravene the G\"udel--Benz relation by
orders of magnitude. This suggests that the chromospheric evaporation model is
not valid for these objects.

\tvlm{}-46546 is an M9 ultra-cool dwarf placed just at the brown dwarf boundary
\citep{2006ApJ...653..690H}. Due to its wide activity spanning from the
radio-domain to the X-rays, \tvlm{} is one of the most intensively studied
ultra-cool dwarfs. The Baraffe models \citep{2003A&A...402..701B} estimate the
mass of \tvlm{} in the range $0.06-0.08$\,M$_{\odot}$, and its radius
$\simeq$\,0.1R$_{\odot}$ for ages older than 0.5\,Gyr.
\citet{2014ApJ...788...23W} estimated the rotation period $\simeq1.96$\,hr using
techniques similar to pulsar timing. \tvlm{} exhibits a variable H$\alpha$
emission and a lack of Li at 670.8\,nm \citep{2002AJ....124..519R}. The
H$\alpha$ emission changes moderately with time, indicating some chromospheric
activity. The observed radio emission suggests a multipolar magnetic field, with
the strength as high as 3\,kG \citep{2006ApJ...653..690H}. \tvlm{} is also the
first ultra-cool dwarf detected with the use of VLBI technique.
\citet{Forbrich2009} observed this object with the VLBA at 8.5\,GHz using the
inner seven stations. They recorded unresolved emission from \tvlm{}. With the
higher spatial resolution allowed by the whole VLBA network, the source appears
marginally resolved with a low signal-to-noise ratio. 

\citet{Forbrich2009} concluded that \tvlm{} could be a binary system with the
projected separation of $\sim$1\,mas. Motivated by first VLBI observations of
\tvlm{}, \citet{Forbrich2013} performed an astrometric survey for a sub-stellar
companion using long-term VLBA observations in 2010 and 2011. The survey
consisted of eight epochs and radio images of \tvlm{} were detected six times.
Based on these observations, \citet{Forbrich2013} excluded companions more
massive than $3.8\,M_\idm{Jup}$ at 16-days orbit and $0.3\,M_\idm{Jup}$ with the
orbital period of 710 days, respectively. The existence of a putative companion
to \tvlm{} is yet uncertain. Recently, \cite{Leto2016} attributed the observed
periodic radio flares in \tvlm{} \citep{2006ApJ...653..690H} to the auroral
emission, analogous to the system of Jupiter and Io. In their model, the $\sim
2$~hr--cyclic pulses may be explained by the presence of a planet orbiting
within the \tvlm{} magnetosphere at short ($<24$~hr) orbital periods, roughly
within 5--15~radii of the object. This hypothesis reinforces the motivation for
further, systematic radio-- and high-precision astrometric monitoring of
\tvlm{}.

%
\section{New EVN observations and data reduction}
%

We made standard continuum observations of \tvlm{} centered at 4.99\,GHz during
seven observational epochs. Stations from Effelsberg, Jodrell Bank (MkII),
Medicina, Noto, Onsala, Toru\'n and Westerbork (phased array) participated in
our observations (observation codes EG053, EG065D, EG065E, EG082A, EG082D \&
EG082E, see Tab.~\ref{tab:tab1} for details). We used the EVN in e-VLBI mode of
observations using the phase-referencing technique. The data were recorded at
1~Gb/s rate, providing the total bandwidth of 128\,MHz divided into 8 base-band
channels with the bandwidth of 16\,MHz each. The first epoch was separated for
two parts due to limited time allocation and both segments were considered
during the reduction as separated epochs. Switching cycle of 5\,min was used
(3.5\,min for target and 1.5\,min for phase calibrator J1455+2131). We also
observed secondary calibrator J1504+2218 applying the same switching cycle.
J1504+2218 is a flat--spectrum compact source located near \tvlm{}, which was
selected from the CLASS
survey\footnote{\url{www.jb.man.ac.uk/research/gravlens/class/class.html}}.
\tvlm{} and J1455+2131 are separated by 1$\degr$.8, \tvlm{} and J1504+2218 by
0$\degr$.9, J1455+2131 and J1504+2218 by 2$\degr$.2, respectively. During
sessions EG053, EG065D \& EG065E, two switching cycles for J1504+2218
observations were added: before and after observations of \tvlm{}. In the case
of EG082, one J1504+2218 switching cycle was added at the beginning of
observations and then it was repeated every five \tvlm{} switching cycles. This
approach resulted in comparable integration time for J1504+2218 during all
experiments but better \emph{uv}--plane coverage. J1504+2218 was observed to
examine the phase--referencing success. We also measured the position of
J1504+2218 at each epoch of scheduled observations.  
\begin{figure*}
\centerline{
  \includegraphics[width=7.6cm]{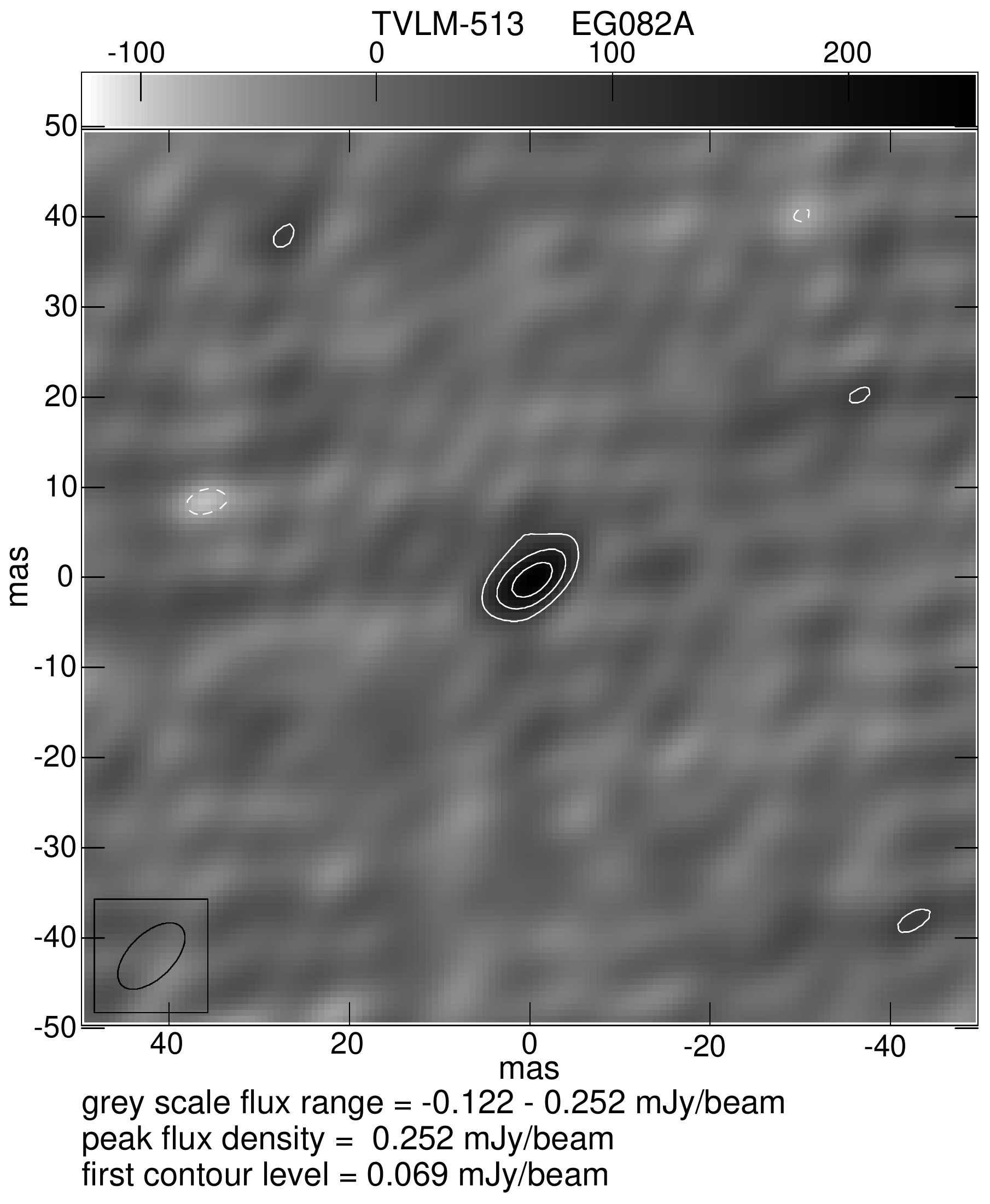}
  \quad
  \includegraphics[width=7.6cm]{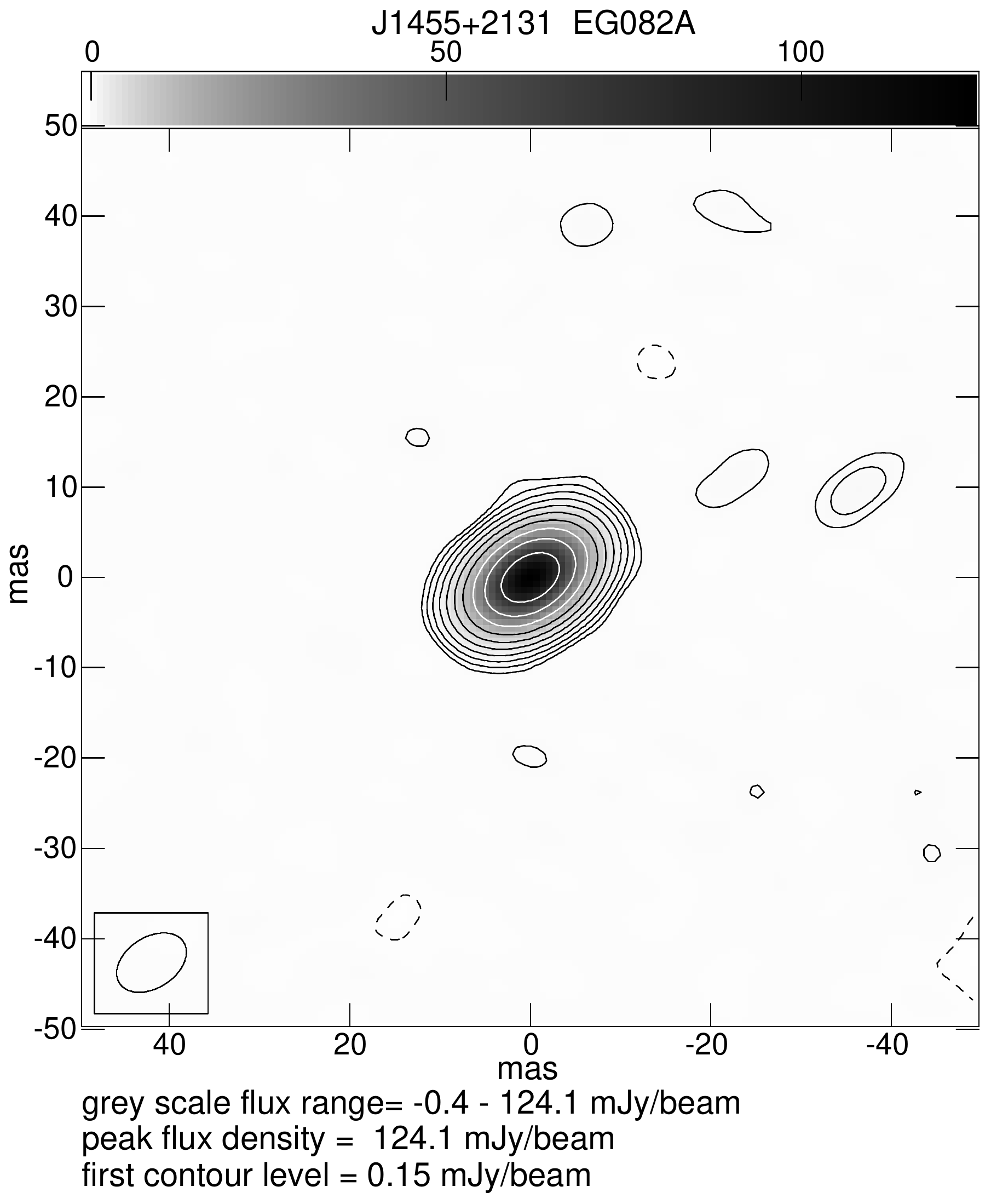}
}
\centerline{
  \includegraphics[width=7.6cm]{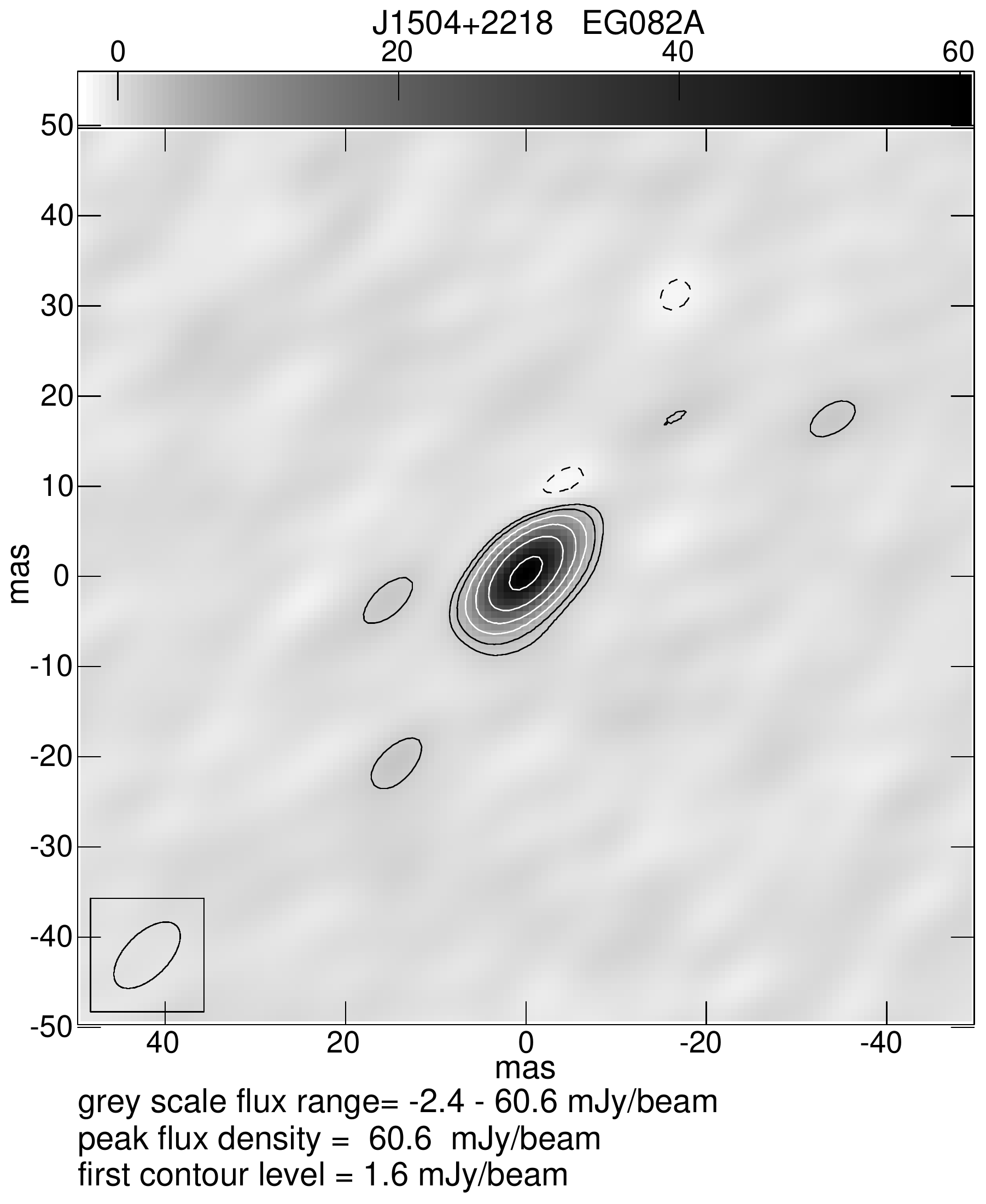}
  \quad
  \includegraphics[width=7.6cm,trim=0 -2.5cm 0 0]{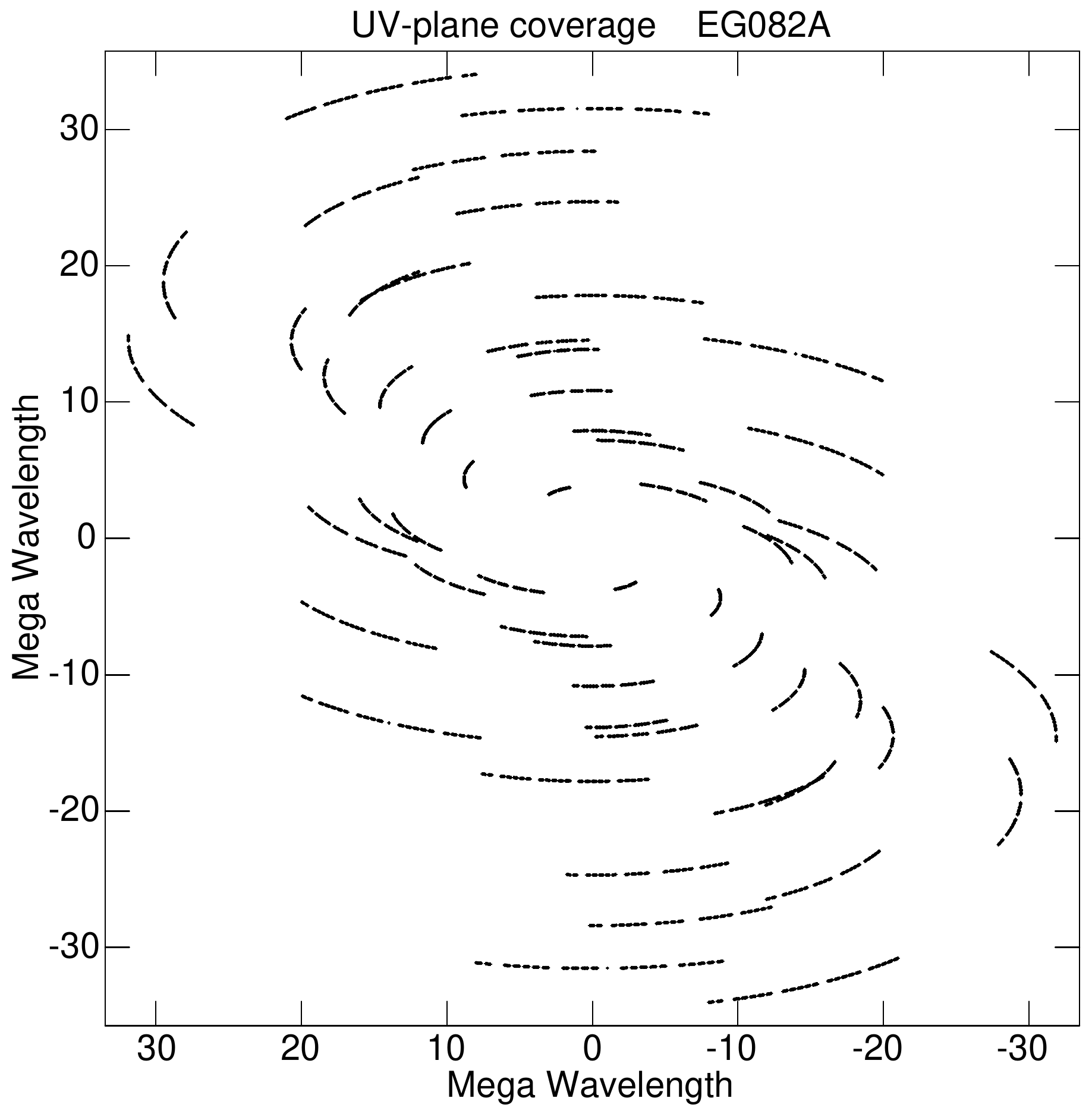}
}
\caption{
Examples of detections in our observational campaign. \emph{Top left}: \tvlm{},
\emph{top right}: J1455+2131, \emph{bottom left}: J1504+2218. All presented
radio maps are based on the data derived during epoch EG082A. The first contour
corresponds to the detection limit of $\simeq\,3\sigma$. Subsequent contour
levels are multiplied by a factor of 2. The in-sets show the size of the
restoring beam. \emph{Bottom right}: a typical \tvlm{} observation
\emph{uv}-plane coverage (also for EG082A).
}
\label{fig:fig1}
\end{figure*}

\begin{table*}
\begin{center}
\begin{tabular}{c c c c c r}
\hline
\hline
Project  & \multicolumn{2}{c}{Date} &  Epoch    & \multicolumn{2}{c}{Conv beam}  \\
 code    & [day] & [UTC]  & (JD-2450000) & [mas] & [deg] \\
\noalign{\smallskip}
\hline
 EG053a & 2011 Mar 11   &  04:07--05:29  &  5631.7001  &  9.4$\times$7.4 & -70~  \\
 EG053b & 2011 Mar 12   &  03:53--05:14  &  5632.6899  &  10.2$\times$9.2  & 26~  \\   
 EG065D & 2012 Oct 9    &  13:11--14:20  &  6210.0705  &  8.7$\times$5.7  & -59~  \\
 EG065E & 2012 Nov 14   &  07:51--09:25  &  6245.8597  &  9.6$\times$5.3 & -41~  \\
 EG082A & 2013 Dec 4    &  05:29--07:37  &  6630.7729  &  9.2$\times$4.9 & -46~  \\   
 EG082D & 2014 Jun 24   &  21:28--23:42  &  6823.4410  &  11.7$\times$5.2 &  50~  \\       
 EG082E & 2015 Mar 25   &  06:05--08:28  &  7106.8031  &  15.8$\times$4.6 &  64~ \\   
\hline
\hline  
\end{tabular}
\end{center}  
\caption[]{The observational log of our astrometric survey of \tvlm{}.}
\label{tab:tab1}
\end{table*} 

\begin{table*}
\begin{center}
\begin{tabular}{c c c c c c c}
\hline
\hline
Project &  \multicolumn{4}{c}{\tvlm{} position}  & \tvlm{} &  J1504+2218\\
 code &  $\alpha$\,(J2000)  & $\Delta$\,$\alpha$ [mas] & $\delta$\,(J2000)  &  
$\Delta$\,$\delta$ [mas] & $S_{5 {\rm GHz}}$ [$\mu$Jy] & $S_{5 {\rm GHz}}$ [mJy] \\
\noalign{\smallskip}
\hline
EG053a &  15 01 08.157048  & 0.63  &  22 50 01.42752 & 0.45  & 238$\pm$51 & 66.5$\pm$1.3 \\
EG053b &  15 01 08.157015  & 0.35  &  22 50 01.42950 & 0.40  & 318$\pm$47 & ---  \\
EG065D &  15 01 08.143333  & 0.23  &  22 50 01.26789 & 0.18  & 331$\pm$41 & 41.5$\pm$1.4 \\
EG065E &  15 01 08.146844  & 0.32  &  22 50 01.24179 & 0.32  & 360$\pm$45 & 52.2$\pm$1.3 \\
EG082A &  15 01 08.145671  & 0.30  &  22 50 01.17079 & 0.27  & 269$\pm$41 & 61.3$\pm$0.8 \\
EG082D &  15 01 08.136623  & 0.47  &  22 50 01.25355 & 0.40  & 227$\pm$42 & 58.4$\pm$0.7 \\
EG082E &  15 01 08.143284  & 0.57  &  22 50 01.17645 & 0.35  & 226$\pm$37 & 41.8$\pm$0.8 \\
\hline
\hline  
\end{tabular}
\end{center}  
\caption[]{
Astrometric position measurements and radio fluxes of \tvlm{} and J1504+2218
collected during the survey. Astrometric uncertainties are formal errors of the
AIPS best-fitting target's positions and do not include any systematic effects.
}  
\label{tab:tab2}
\end{table*} 
The data reduction process was carried out using the standard NRAO
AIPS\footnote{\url{www.aips.nrao.edu/index.shtml}} procedures. The visibility
data were inspected for quality, and noisy points were removed using AIPS task
EDITR. Maps of the phase calibrator J1455+2131 were created separately for each
epoch and were used as a reference model for the final fringe--fitting. Before
the fringe--fitting, we corrected ionospheric Faraday rotation and dispersive
delay using AIPS task TECOR. The IMAGR task was used to produce the final total
intensity images of all our sources. Radio fluxes and estimated positions of
\tvlm{} and J1504+2218 were then measured by fitting Gaussian models, using AIPS
task JMFIT and are presented in Tab.~\ref{tab:tab2}. In order to study the
radio-emission variability of \tvlm{}, we reconstructed its light curves using
AIPS task DFTPL. Before the application of DFTPL, we searched for background
objects within $3\arcsec \times 3\arcsec$ around \tvlm{} position and none
were found. If a background object would be detected then the resulting source
model should be subtracted from the visibility data. Its side-lobes and shape
changes of the synthesized beam could result in flux variations over the radio
map and they might ``contaminate'' the real variability or even may generate a
false signal.
\begin{figure*}
\centerline{
    \hbox{\includegraphics[width=0.32\textwidth]{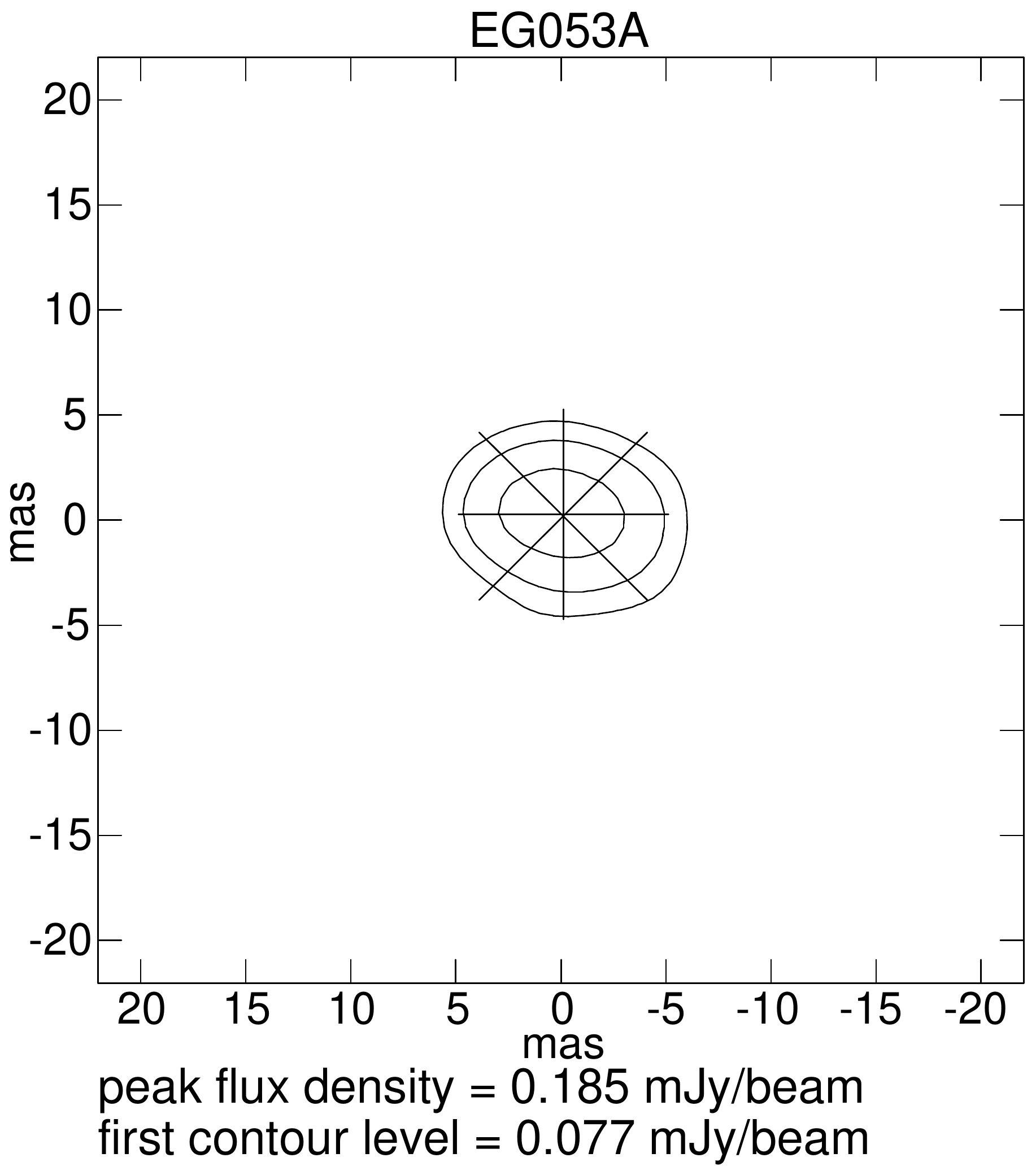}}
    \quad
    \hbox{\includegraphics[width=0.32\textwidth]{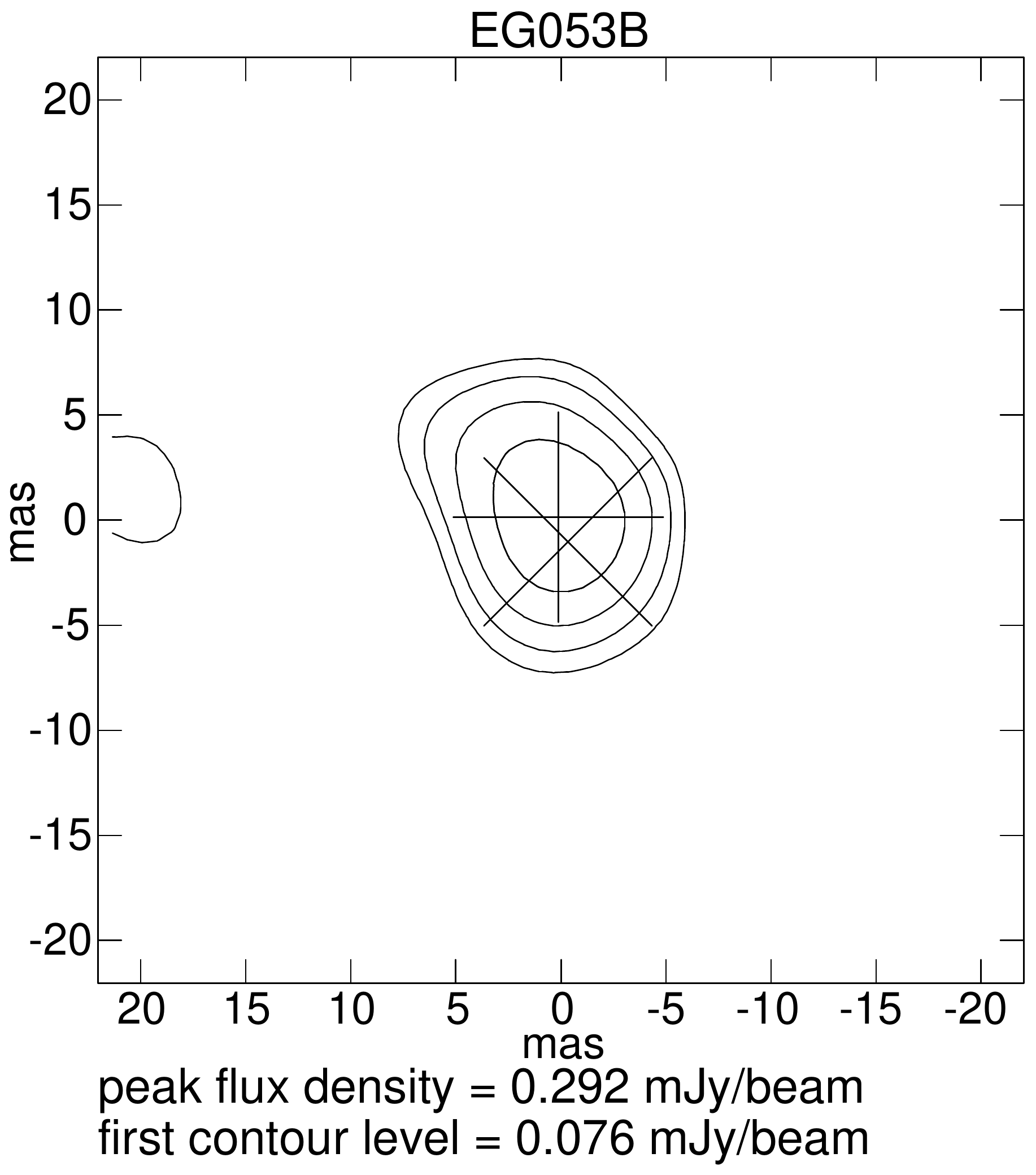}}
    \quad
    \hbox{\includegraphics[width=0.32\textwidth]{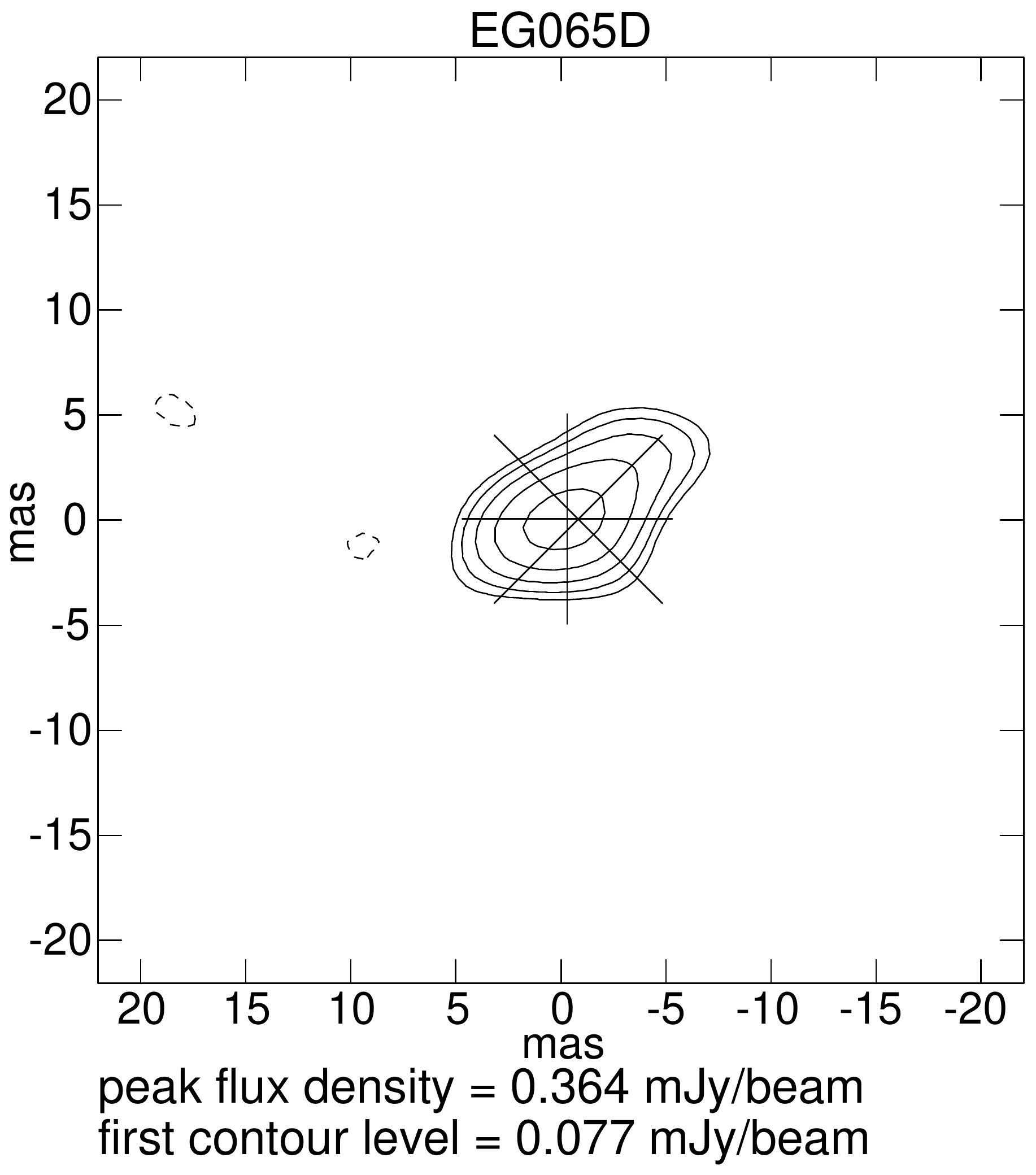}}
    }
\medskip
\centerline{
    \hbox{\includegraphics[width=0.32\textwidth]{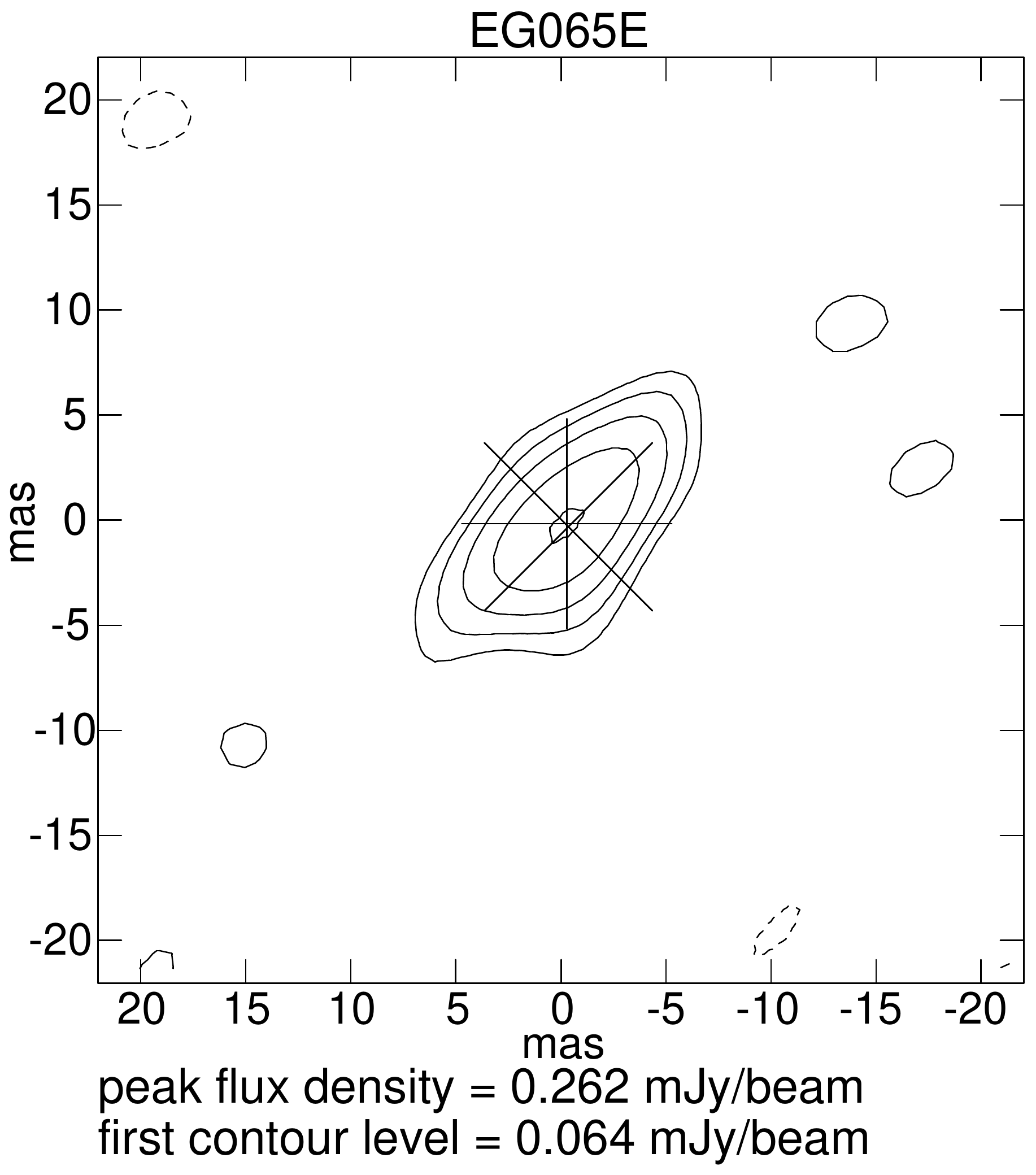}}
    \quad
    \hbox{\includegraphics[width=0.32\textwidth]{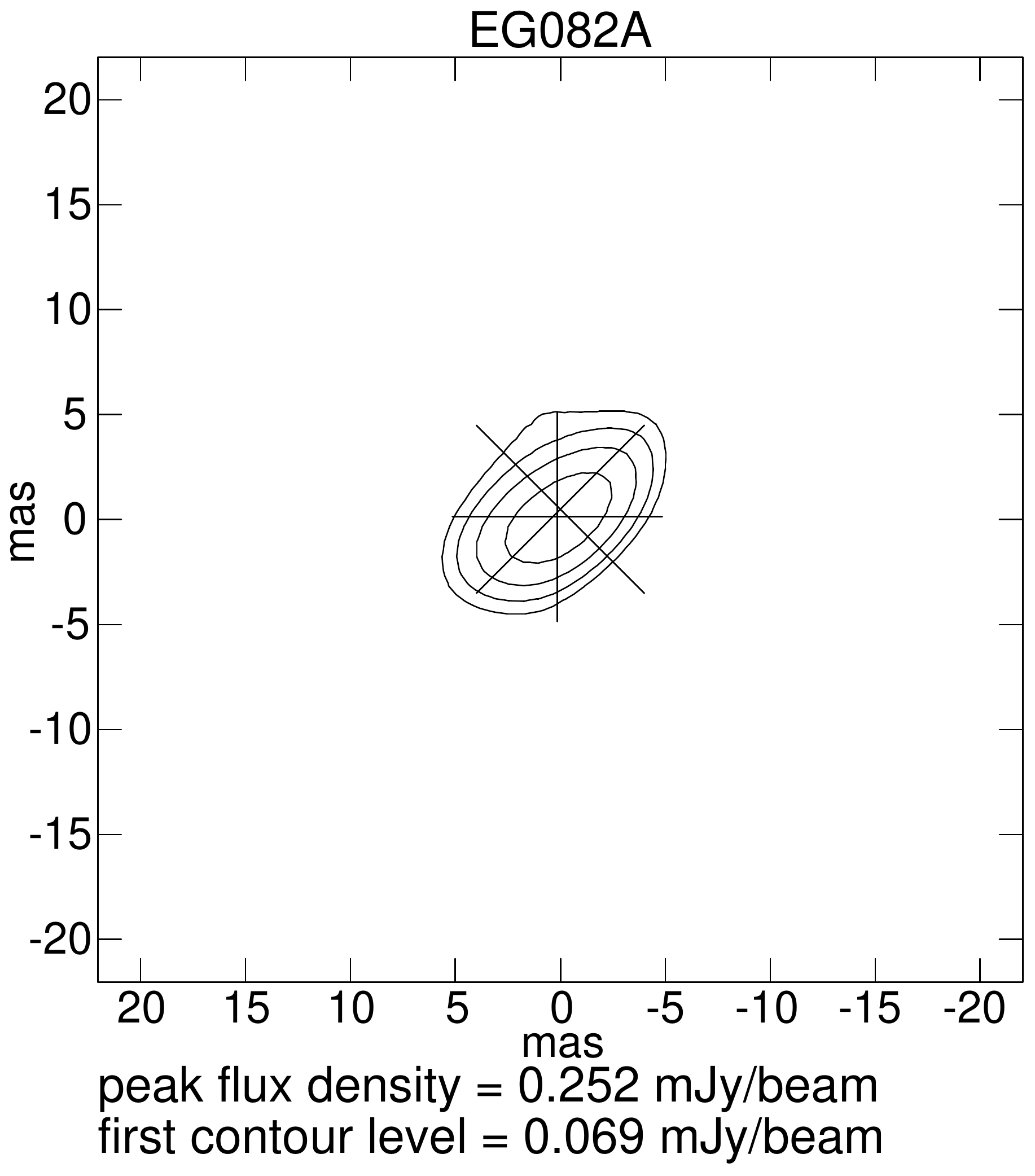}}
    \quad
    \hbox{\includegraphics[width=0.32\textwidth]{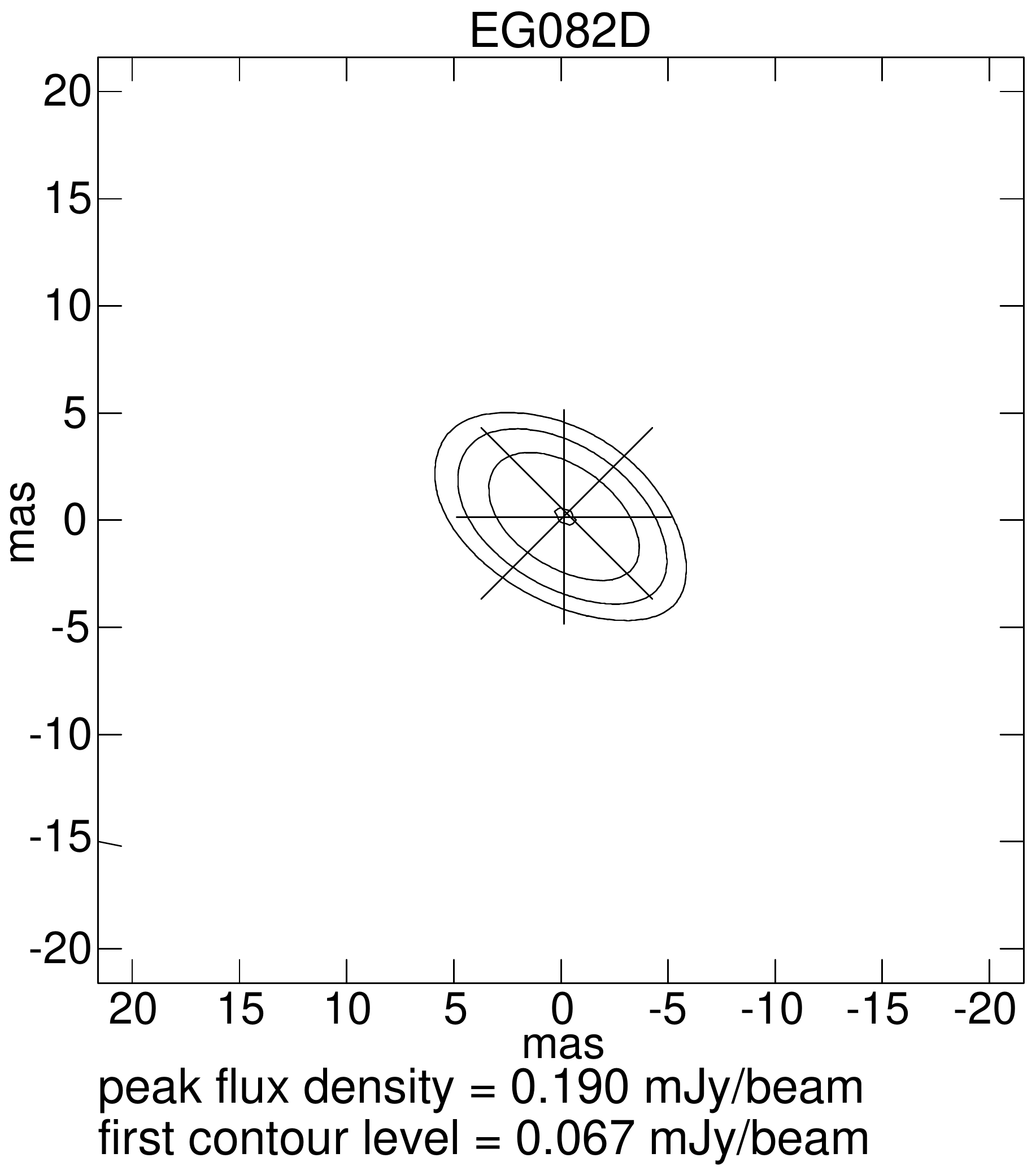}}
    }
\caption{
Radio maps of \tvlm{} based on data collected during RISARD project. The first
contour corresponds to the detection limit of $\simeq\,3\sigma$. Successive
contour levels are multiplied by a factor of 2. The $+$ symbols mark the
measured positions, and the $\times$ symbols represent the model astrometric
position at the particular epoch of observations, respectively.
}
\label{fig:fig2}
\end{figure*}

\begin{figure}
\centerline{
    \hbox{\includegraphics[width=0.42\textwidth]{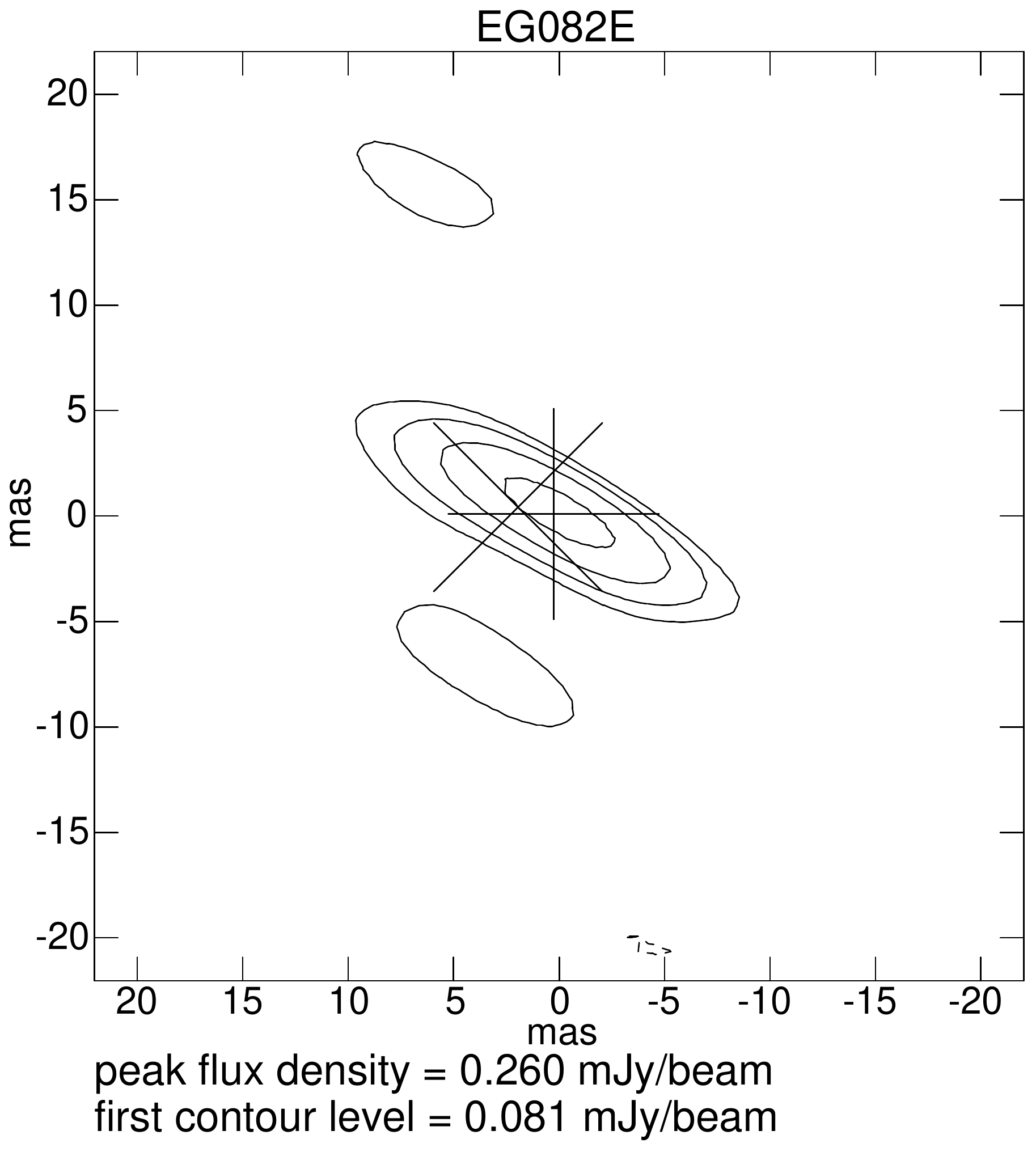}}
    }
\label{fig:fig3}
\caption{
Radio maps of \tvlm{} based on the observational data collected during RISARD
project. The first contour corresponds to the detection limit of
$\simeq\,3\sigma$. Successive contour levels are multiplied by a factor of 2.
The $+$ symbol marks the measured position and the $\times$ symbol
represents the model position at the epoch of observations, respectively.
}
\end{figure}

Understanding sources of uncertainties of astrometric VLBI observations is
crucial for their correct estimation. There are a few origins of systematic
errors, like the residual phase in phase-referencing, sub-mas changes of the
phase calibrator structures, or differences between optical path lengths for the
target and the phase calibrator. The later are caused by the atmospheric zenith
delay residuals. There are mainly two different solutions of this problem.
Assuming uncorrelated and normal uncertainties, one may introduce a'priori
introduced ``error floor'' added in quadrature to raw uncertainties in order to
obtain the reduced $\chi_{\nu}^2\simeq 1$
\citep[e.g.,][]{2014PASJ...66..104,Forbrich2013}. Additional interferometric
observations of compact extragalactic sources spread over the sky are used to
measure broad-band delays \citep{2009ApJ...693..397R}. It should be also
mentioned that properties of M-dwarf radio emission could result in further
scatter of astrometric measurements. \citet{1998A&A...331..596B} showed that the
radio corona of an active M-dwarf UV\,Cet\,B varies in size and position during
a large radio flare. However, the impact of flaring events on the radio
astrometry could be reduced if it is possible to remove such events from the
analysis and the quiescent emission is strong enough to obtain radio images with
reasonable SNR$\gtrsim$5. We detected strong $\sim$1\,mJy, circularly polarized
flares which occurred during our observations (see Sect. 5 for details).
These events spanned short time intervals ($t\lesssim3$\,min) in comparison with
the duration of observations ($\sim2$\,hrs of integration time per epoch).
Therefore we decided to not remove the identified flares from the mapping process.

Here we present a different, systematic statistical approach to the proper
optimization of astrometric model, Eqs.~\ref{eq:m1}--\ref{eq:m2}, in the
presence of unspecified error factors. It is based on the maximal likelihood
function and Markov Chain Monte Carlo (MCMC) exploration of the parameters
space. The error floor estimated in this way accounts for different systematic
effects, spanning atmospheric phase effects, a possible binarity of the target,
and the motion of unseen, low-mass companions.

%
\section{Astrometric model and its  optimization}
%

To determine the parallax and components of the proper motion, we apply
a general, 7-element astrometric model with the secular acceleration terms:
\begin{eqnarray}
 \alpha(t_i) &=& \alpha_0 + \mu_\alpha ( t_i - t_0 ) 
            + \pi_\alpha(t_i, \alpha,\delta) + a_\alpha(t_i - t_0)^2, 
\label{eq:m1}   \\               
 \delta(t_i) &=& \delta_0 + \mu_\delta ( t_i - t_0 ) 
            + \pi_\delta(t_i, \alpha,\delta) + a_\delta(t_i - t_0)^2,
\label{eq:m2} 
\end{eqnarray}
where ($\alpha_0,\delta_0$) are the target's ICRF coordinates and
$(\mu_\alpha,\mu_\delta)$ are components of the proper motion at the initial
epoch $t_0$, respectively; $(\pi_\alpha,\pi_\delta)$ are the parallax factors
(i.e., the parallax $\pi$ projected onto the coordinates axes), and
$(a_\alpha,a_\delta)$ are components of the secular acceleration relative to the
initial epoch $t_0$. The secular acceleration terms are included to express a
possible long-term perturbation to the inertial motion of the target and/or the
perspective (geometric) acceleration. We considered also 5-elements model
without the acceleration terms. (It will be explained below that in fact our 5-
or 7-parameter models are optimized with an additional parameter scaling
measurements errors).

To get rid of $(\alpha,\mu_\delta)$ and $(\delta,\mu_\delta)$ correlations, we
choose the initial epoch $t_0$ as the mean of all observational epochs $t_i$
weighted by uncertainties $\sigma_i$ ($i=1,\ldots,N$),
\begin{equation}
t_0 = \frac{\sum_i^{N} t_i w_i}{\sum_i^{N} w_i}, \quad
w_i=\frac{1}{\sigma_i}.
\label{eq:t0}
\end{equation}
Since our preliminary fits revealed $\Chi \sim 2$ suggestive for underestimated 
uncertainties, we optimized the maximum likelihood function ${\cal L}$:
\begin{equation}
 \log {\cal L} =  
-\frac{1}{2} \sum_{j,t} \frac{\mbox{(O-C)}_j^2}{{\sigma_j}^2}
- \frac{1}{2}\sum_{j} \log {{\sigma_j}^2} 
- \frac{1}{2} M \log{2\pi},
\label{eq:Lfun}
\end{equation}
where $(\mbox{O-C})_{j,t}$ is the (O-C) deviation of the observed $\alpha(t_i)$
or $\delta(t_i)$ at epoch $t_i$ from its astrometric ephemeris, and their
uncertainties are $\sigma_{j}^2 \rightarrow \sigma_{j}^2+\sigma_f^2$ with a
parameter $\sigma_f$ scaling raw uncertainties (the error floor), and
$j=1\ldots,M$ where $M=2N$ is the total number of $(\alpha,\delta)$
measurements. We assume that uncertainties $\sigma_j$ are Gaussian and
independent. By introducing the scaling of uncertainties, we aim to determine
the error floor in a self-consistent manner, instead of fixing it {\em
a'posteriori}, as in \citet{Forbrich2013}. 

Combined 7 VLBA measurements in \citet{Forbrich2009} and \citet{Forbrich2013}
with our 7 EVN detections result in 28 $(\alpha,\delta)$-datums, spanning
$\Delta t=2550.9222$~days. Given raw uncertainties in this data set, we computed
the initial epoch $t_0=$JD~2455424.19763 in accord with Eq.~\ref{eq:t0}.

We optimized the $\log{\cal L}$ function indirectly with the Markov Chain Monte
Carlo (MCMC) technique. We determine the posterior probability distribution
${\cal P}(\tv{\xi}|{\cal D})$ of astrometric model parameters $\tv{\xi} \equiv
[\alpha_0,\delta_0,\pi,\mu_\alpha,\mu_\delta,a_\alpha,a_\delta,\sigma_f]$ in
Eqs.~\ref{eq:m1}--\ref{eq:m2}, given the data set ${\cal D}$ of all astrometric
observations (understood as $\alpha_i$ and $\delta_i$ components): 
$
  {\cal P}(\tv{\xi}|{\cal D}) 
  \propto {\cal P}(\tv{\xi}) \, {\cal P}({\cal D}|\tv{\xi}),
$
where ${\cal P}(\tv{\xi})$ is the prior, and the sampling data distribution
${\cal P}({\cal D}|\tv{\xi}) \equiv \log{\cal L}(\tv{\xi},{\cal D})$. For all
parameters, besides the acceleration terms, we define priors as flat (or uniform
improper) by placing limits on model parameters, i.e., $\alpha_{0}>0$,
$\delta_{0}>0$, $\mu_{\alpha}>0$, $\mu_{\delta}>0$ $\pi>0$ and $\sigma_f>0$. For
the acceleration terms, which magnitude is unspecified, we applied the Jeffreys
prior:
\[
  {\cal P}(\xi) = \frac{1}{\xi_\idm{min}+\xi},
\]
where $\xi_\idm{min}$ is a small value to avoid underflows.

To perform the MCMC sampling of the posterior, we used the affine-invariant ensemble MCMC sampler
\citep{Goodman2010} encoded in a great {\sc emcee} package invented and
developed by \citet{Foreman2014}. To compute the parallax factors, we used the
{\sc DE405} ephemeris and subroutines from the {\sc NOVAS} package
\citep{Kaplan2010}.

\begin{figure*}
\centerline{
    \hbox{\includegraphics[width=0.862\textwidth]{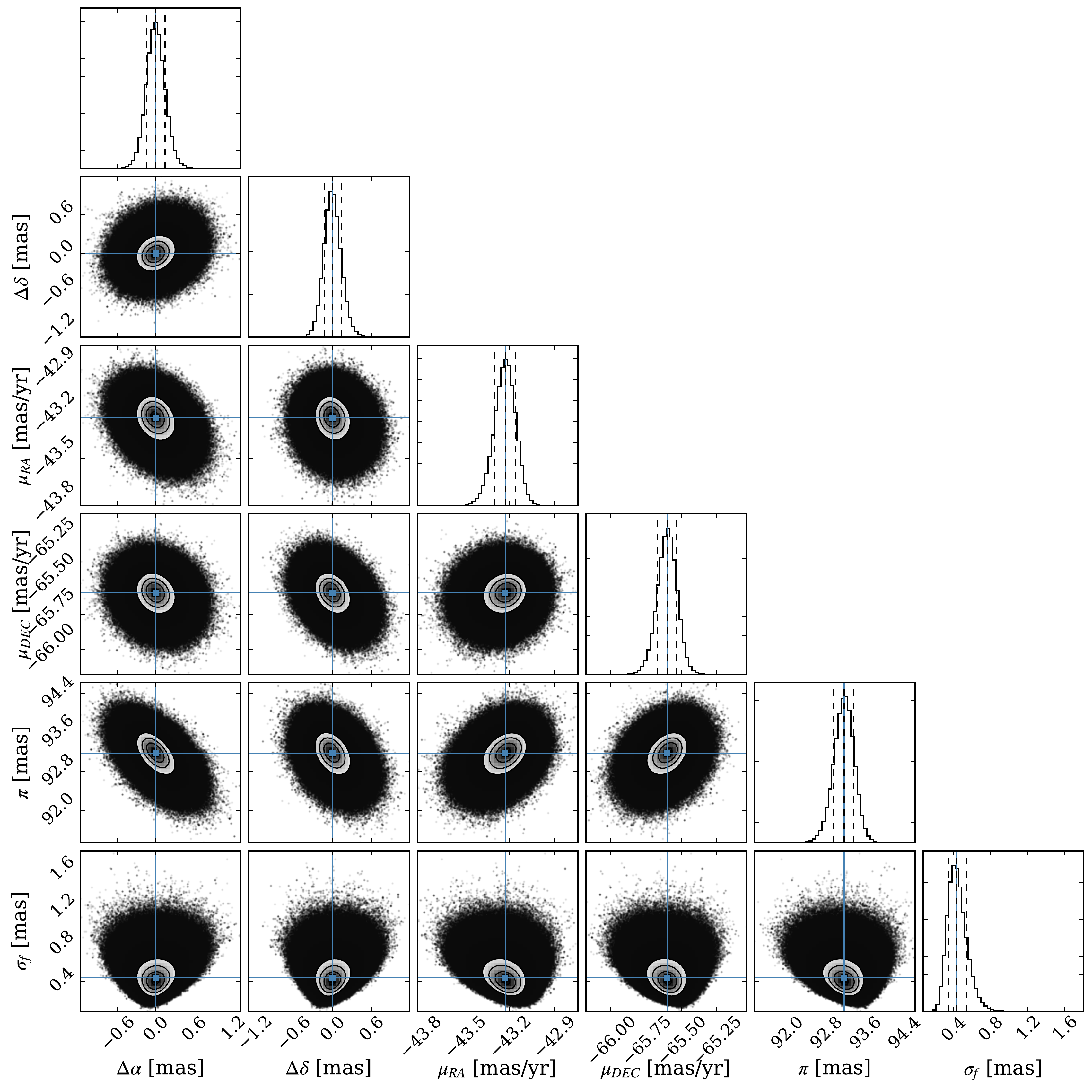}}
    }
\caption{
One-- and two--dimensional projections of the posterior probability distribution
for all free parameters of the astrometric 5-elements model. The MCMC chain
length is 256,000 iterations in each of 640 different initial conditions
in a small ball around a preliminary astrometric model derived with a common
function minimization. Contours indicate 16th, 50th, and 84th percentiles of
the samples in the posterior distributions. Crossed lines illustrate the
best-fitting parameters displayed in Tab.~\ref{tab:tab3}.
}    
\label{fig:fig4}
\end{figure*}
\begin{figure}
\centerline{
    \hbox{\includegraphics[width=0.5\textwidth]{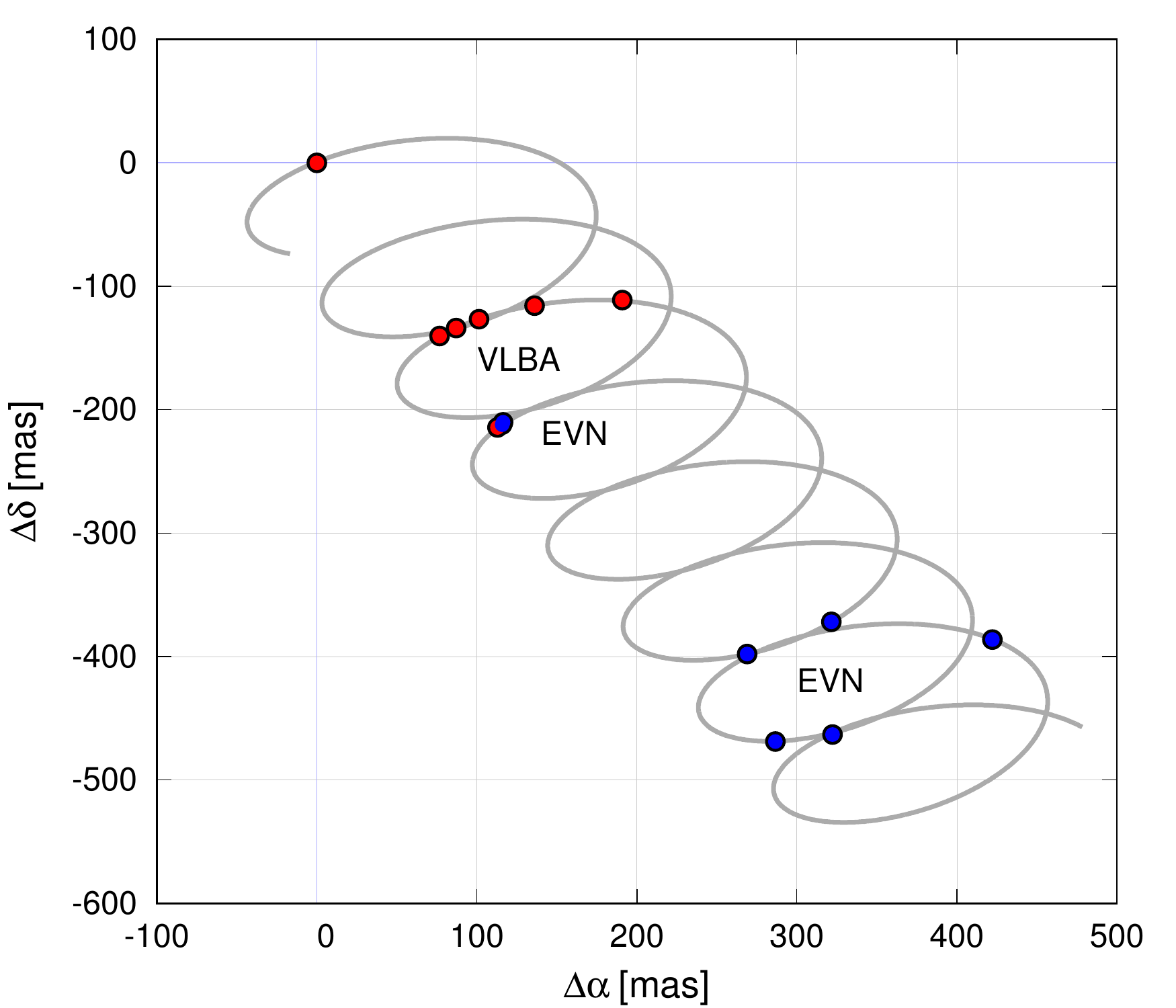}}
    }
\caption{
Sky-projected, 5-parameter astrocentric model of \tvlm{} positions (grey curve)
over-plotted with the VLBA (red filled circles) and with all new EVN detections
(blue filled circles, this work), relative to the first observation in
\citep{Forbrich2013}.
}
\label{fig:fig5}
\end{figure}

\begin{figure*}
\centerline{
    \hbox{\includegraphics[width=0.5\textwidth]{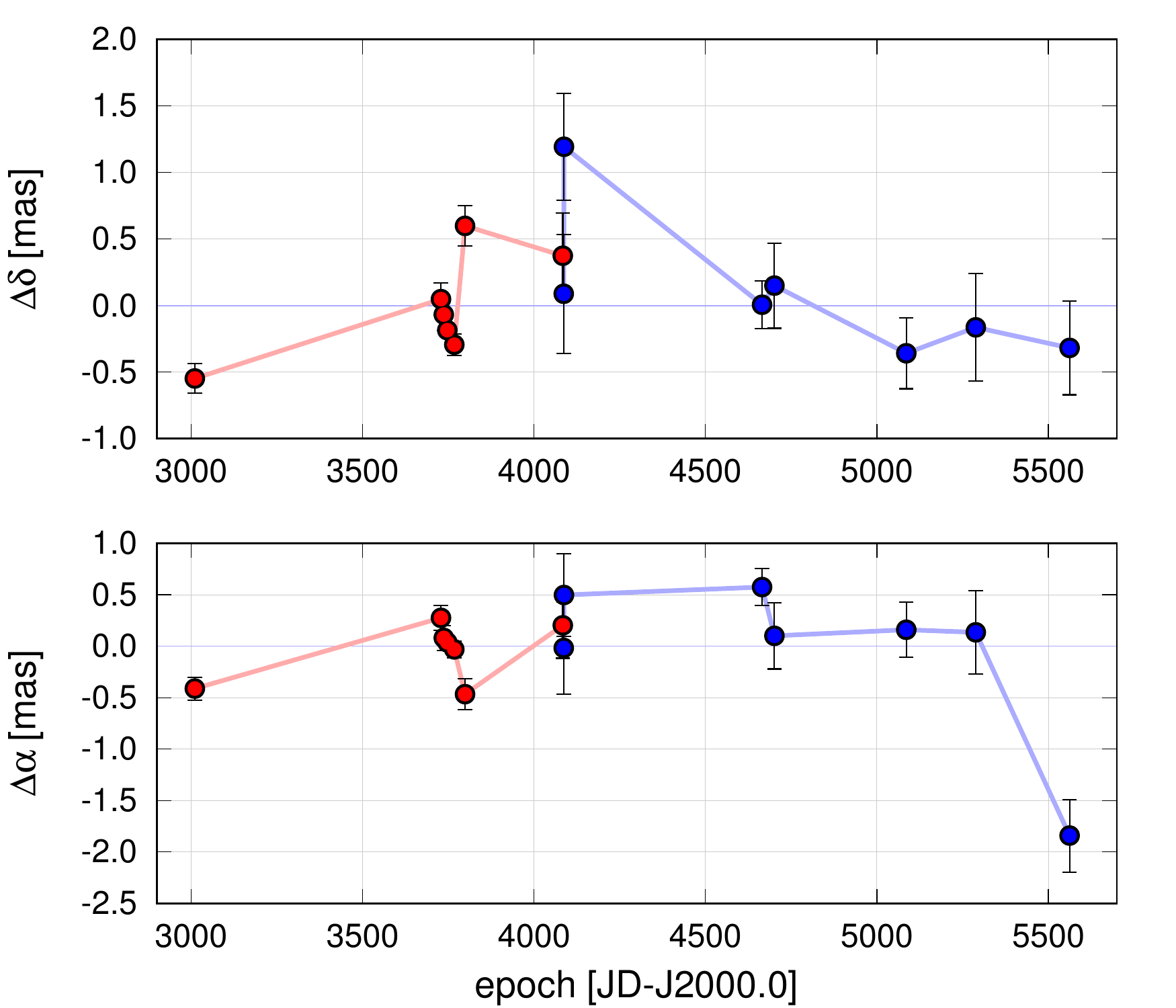}}
    \hbox{\includegraphics[width=0.5\textwidth]{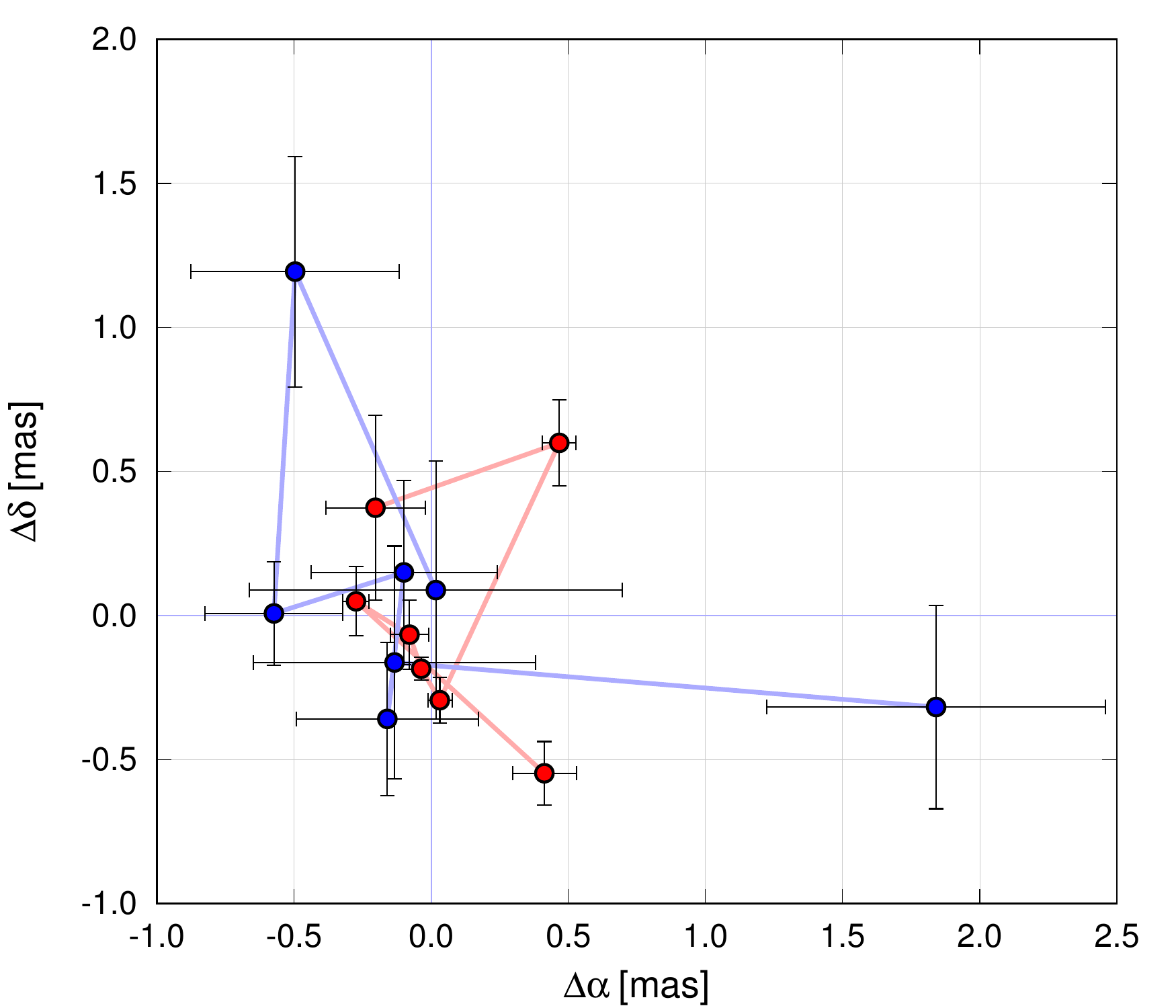}}
   }
\caption{
Residuals in right ascension ($\alpha$) and declination ($\delta$) to
5-parameter astrocentric model (Eqs.~\ref{eq:m1}--\ref{eq:m2} and
Fig.~\ref{fig:fig5}) plotted as a function of the observational epochs ({\em
the left column}). A plot in {\em the right panel} show the residuals in the
coordinates plane. Error bars mark the reported, instrumental uncertainties
which do not include systematic effects. The VLBA measurements are marked with
red open circles and the EVN data (this paper) are marked with blue open
circles.
}   
\label{fig:fig6}
\end{figure*}

We performed a number of experiments by increasing the MCMC chain lengths up to
512,000 samples. The posterior probability distribution for the 5-elements model
is illustrated Fig.~\ref{fig:fig4}. It shows one-- and two-dimensional
projections of the posterior for all free parameters of the model. A well
defined solution is apparent. No significant parameter correlations are present.
The best-fitting parameter values and their uncertainties estimated between the
16th and 86th percentile are displayed in Tab.~\ref{tab:tab3} and the on-sky
motion of the target is illustrated in Fig.~\ref{fig:fig5}. The residuals
to the 5-parameter model are illustrated in Fig.~\ref{fig:fig6}. The
best-fitting solution exhibits the error floor as large as~0.43 mas, which is
roughly two times larger than estimated by \cite{Forbrich2013} for their VLBA
observations alone.
\begin{table}
\begin{center}
\begin{tabular}{l c c}
\hline
\hline
parameter  & 5-element fit        & 7-element fit \\
\hline
\smallskip
$\alpha_0$ & $15^{\idm{h}}01^{\idm{m}}8^{\idm{s}}.15219_{-0.00010}^{+0.00009}$ 
           & $15^{\idm{h}}01^{\idm{m}}8^{\idm{s}}.15219_{-0.00010}^{+0.00009}$
\smallskip\\
$\delta_0$ & $22^{\circ} 50' 1''.42470_{-0.00070}^{+0.00075}$ 
           & $22^{\circ} 50' 1''.42470_{-0.00014}^{+0.00013}$ 
\smallskip\\
$\mu_{\alpha}$ [mas yr$^{-1}$] &  -43.23${}_{-0.07}^{+0.08}$
                               &  -43.13${}_{-0.07}^{+0.08}$ 
\smallskip\\
$\mu_{\delta}$ [mas yr$^{-1}$] & -65.60${}_{-0.07}^{+0.08}$
                               & -65.50${}_{-0.07}^{+0.07}$ 
\smallskip\\
$a_{\alpha}$ [mas cyr$^{-2}$] &  ---
                          &  -12.5${}_{-4.8}^{+5.1}$ 
\smallskip\\
$a_{\delta}$ [mas cyr$^{-2}$] &  ---
                            & -11.3${}_{-4.5}^{+4.5}$ 
\smallskip \\
parallax $\pi$ [mas]     &  93.17${}_{-0.20}^{+0.21}$
                         &  93.27${}_{-0.17}^{+0.18}$
\smallskip \\
$\sigma_f$  [mas]       &   0.43${}_{-0.12}^{+0.10}$ 
                        &   0.36${}_{-0.10}^{+0.08}$
\\
\noalign{\smallskip}
\hline
\hline
\end{tabular}
\end{center}  
\caption[]{
Parameters of the best-fitting solution at the middle-arc epoch
$t_0$=JD~2455424.19763.
}
\label{tab:tab3}
\end{table} 

\begin{figure}
\centerline{
    \hbox{\includegraphics[width=0.5\textwidth]{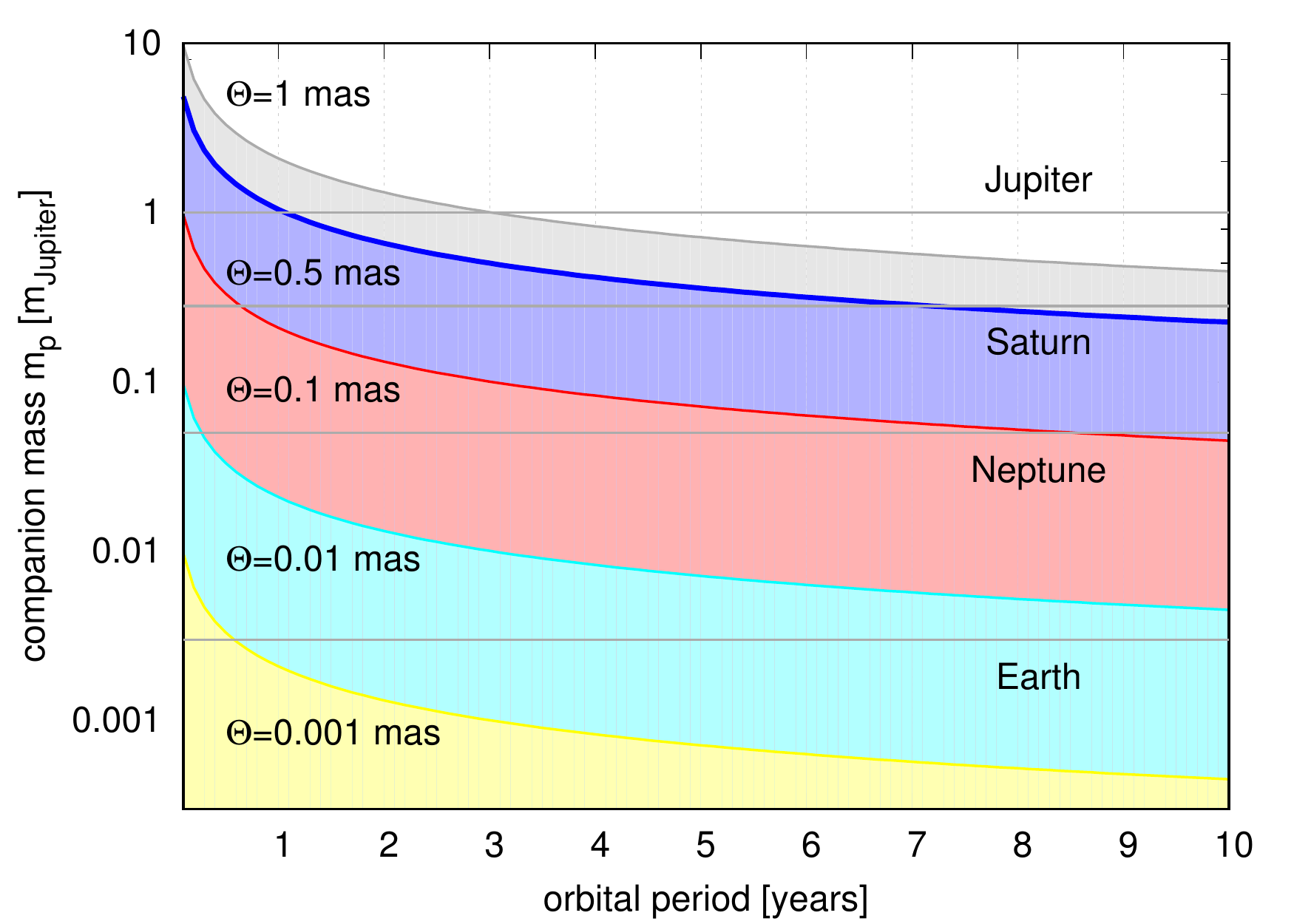}}
    }
\caption{
Astrometric detection limits for a companion to \tvlm{} in the (orbital
period,mass)--space (see Eq.~\ref{eq:mass}) for the mass of \tvlm{}
$m_{\star}=0.08$~M$_{\odot}$ and its derived distance $d=10.733$~pc.
}
\label{fig:fig7}
\end{figure}
Given the astrometric residuals to the 5-parameter model, we may estimate the
mass range of a hypothetical companion, which could be present below the
detection limit. Assuming that such a companion exists in a circular
Keplerian orbit with semi-major axis~$a$, orbital period~$P_\idm{p}$, and
mass~$m_{\idm{p}}$, such a body would cause the reflex motion of the target
around the barycenter with an angular semi-amplitude of:
\begin{equation}
 \Theta = \frac{m_{\idm{p}}}{m_{\star}} 
 \left[ \frac{P_\idm{p}^2}{4\pi^2} k^2 (m_{\idm{p}}+m_{\star})
\right]^{-1/3}, 
\label{eq:mass}
\end{equation}
where $m_{\star}$ and $m_{\idm{p}}$ are the masses of the star and its
companion, respectively, $k^2$ is the Gauss gravitational constant and
$P_{\idm{p}}$ is the orbital period of the companion, see also
\citet{Forbrich2013}. For a known or assumed mass of the binary and given its
orbital period and angular separation $\Theta$ from the primary, this relation
may be solved w.r.t. $m_{\idm{p}}$. Parametric plots $m{\idm{p}}\equiv
m_{\idm{p}}(P_{\idm{p}},\Theta)$ for a few borderline angular separations are
shown in Fig.~\ref{fig:fig7}. Corresponding mass detection levels for a few
characteristic objects are labelled. 

Unfortunately, the sampling and relatively low astrometric accuracy of our EVN
measurements does not make it possible to resolve any clear, systematic reflex
motion of the primary (the right panel of Fig.~\ref{fig:fig6}). 

Moreover, astrometric positions at two epochs of EG053b and EG082E
(Fig.~\ref{fig:fig3}) deviate by $\gtrsim 1$--$2\,$mas from the 5-parameter
model. These excessively large discrepancies could be most likely explained
through pure observational and local effects. During EG053b (12-th of March,
2011) a strong geomagnetic storm was present in the Earth ionosphere and the
aurora was visible all above Europe. At the EG082E epoch, \tvlm{} elevation for
the European station was low (20$^{\circ}$--30$^{\circ}$ above the horizon) and
sparse $uv$--coverage resulted in a large, elongated convolution beam. These
conditions of observations lead to extensive phase errors that result in shifted
and non-Gaussian radio images of the star. That is especially important in the
case of elongated convolution beams. 

Yet we attempted to model a potential curvature effect with 7-parameter model,
Eq.~\ref{eq:m1} \& \ref{eq:m2}. The results are shown in the right-hand column
of Tab.~\ref{tab:tab3}. At this time, the error floor is slightly smaller than
for the 5-parameter model, and the curvature coefficients are roughly
$-12$~mas/cy$^{-2}$. Such large values might indicate a massive companion and/or
a significant perspective acceleration. However, given the apparent curvature is
caused by two strongly outlying measurements (Fig.~\ref{fig:fig6}), we found 
that the residuals actually vary within $\sim 0.5$~mas, when centered at the
$t_0$ epoch position. Therefore we may rule-out companions more massive than
Saturn in $\gtrapprox7$--yr orbit or Jupiter in $\gtrapprox1$~yr orbit. However,
very short-period companions (roughly below $\sim 1$~year time-scale) within
mass limits illustrated in Fig.~\ref{fig:fig7} cannot be excluded due to sparse
sampling. Our estimates are consistent with the results of \citet{Forbrich2013}.
Unfortunately, we cannot confirm nor rule their hypothesis of short-period
companion in $\sim 16$~days orbit, as well as a putative close-in,
short-period planet triggering periodic auroral activity due to the interaction
with the magnetosphere of \tvlm{} \citep{Leto2016}. 
We note that such an explanation of the observed
periodic radio pulses from low-mass stars has been originally proposed by 
\cite{2015Natur.523..568H}. Regrettably, the astrometric observations 
in \cite{Forbrich2013} and in this work are currently not 
enough sensitive to detect such short-period planets.
%
\section{Properties of observed radio emission}
%

Since we observed \tvlm{} across a wide time-window of 4 years (between March
2011 to March 2015), the EVN data make it possible to track the radio
variability for over short (a~few minutes) to long (a~few months) time scales. The
radio emission traces particle acceleration by magnetic field in stellar
coronae, and corresponds to incoherent (gyrosynchrotron emission) or coherent
radiation (electron cyclotron maser or plasma emission). 

\begin{figure*}
\centerline{
  \includegraphics[angle=0,width=8.3cm]{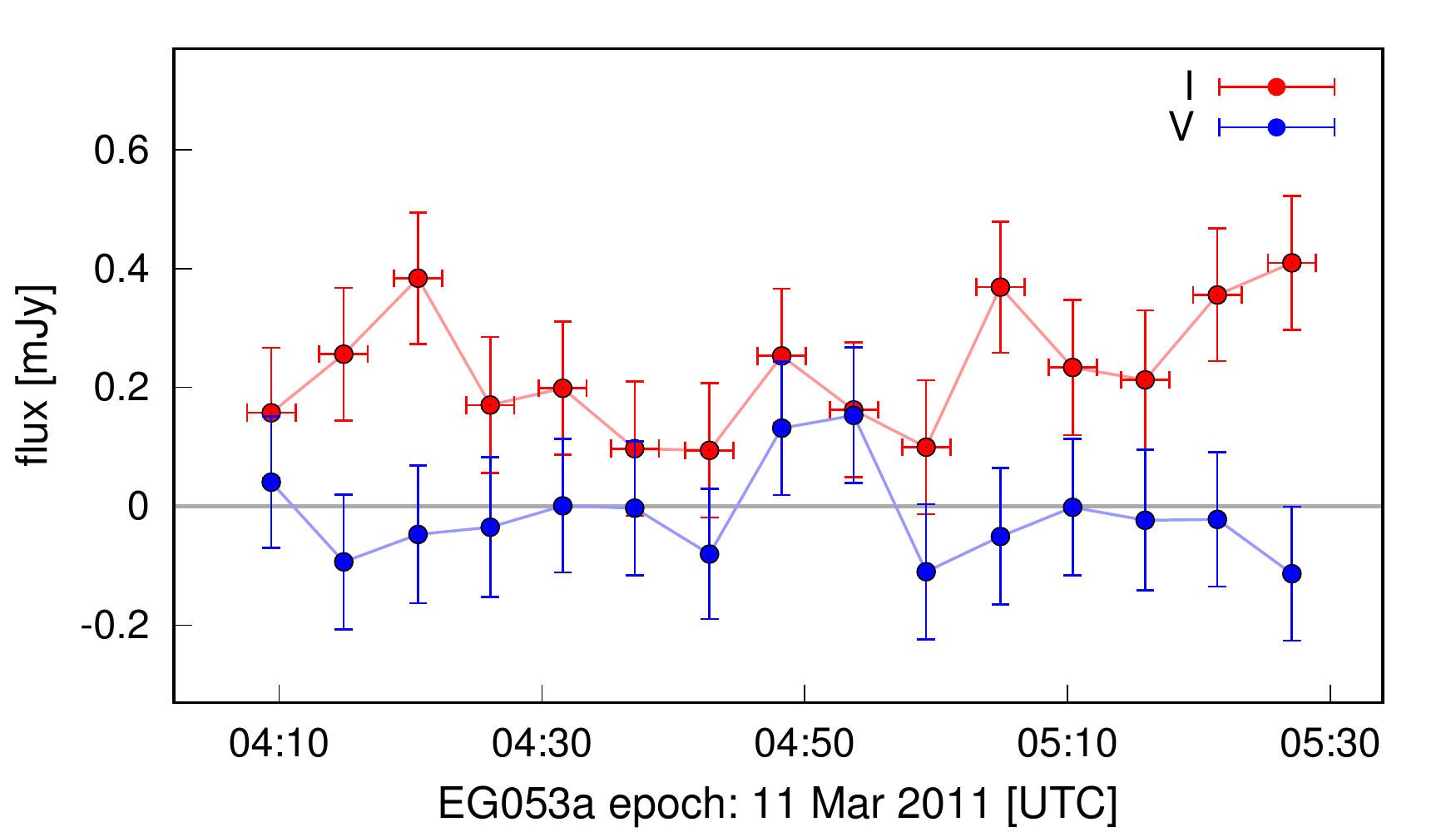}
  \includegraphics[angle=0,width=8.3cm]{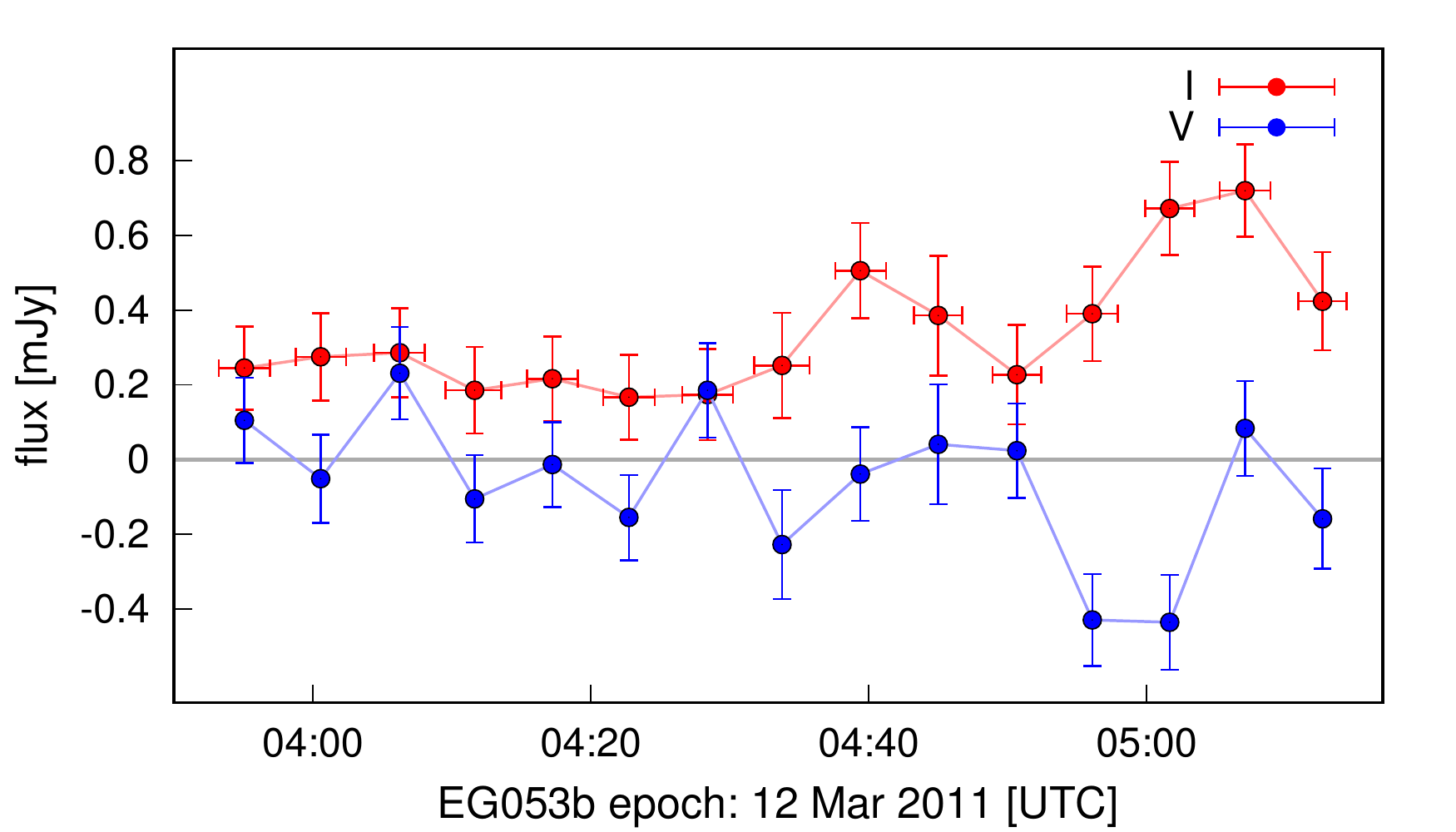}
}
\centerline{
  \includegraphics[angle=0,width=8.3cm]{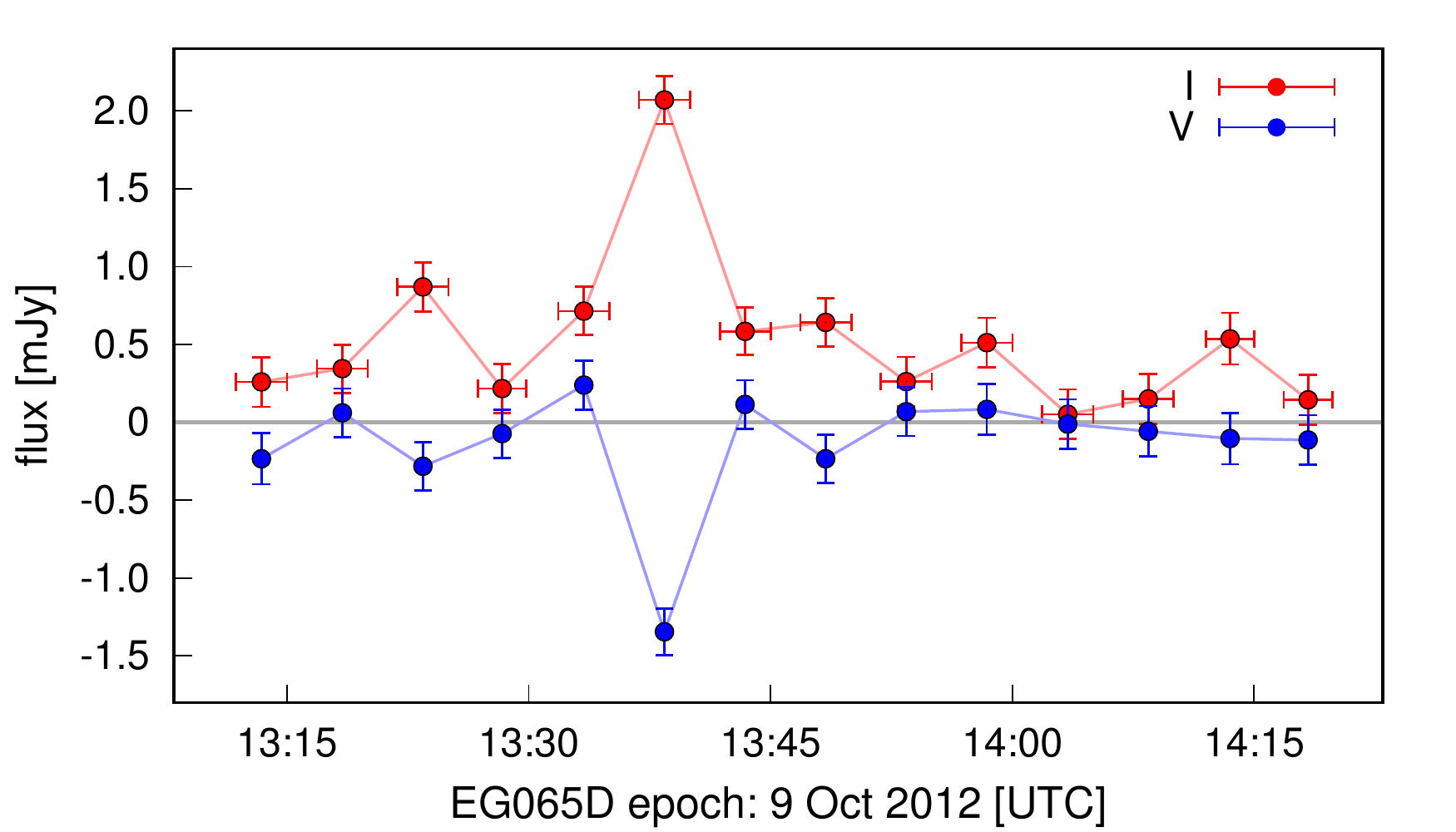}
  \includegraphics[angle=0,width=8.3cm]{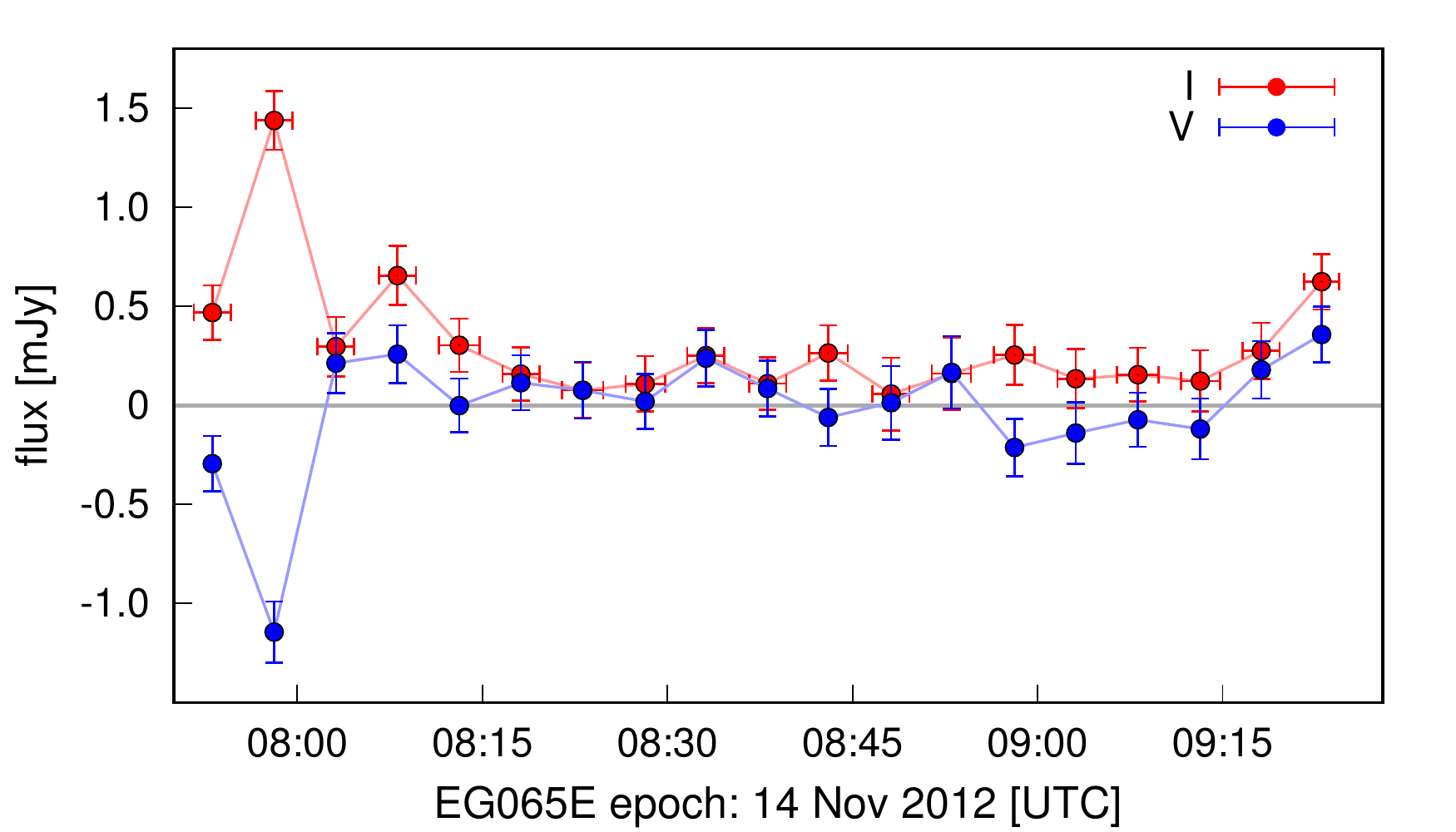}
}
\centerline{
  \includegraphics[angle=0,width=8.3cm]{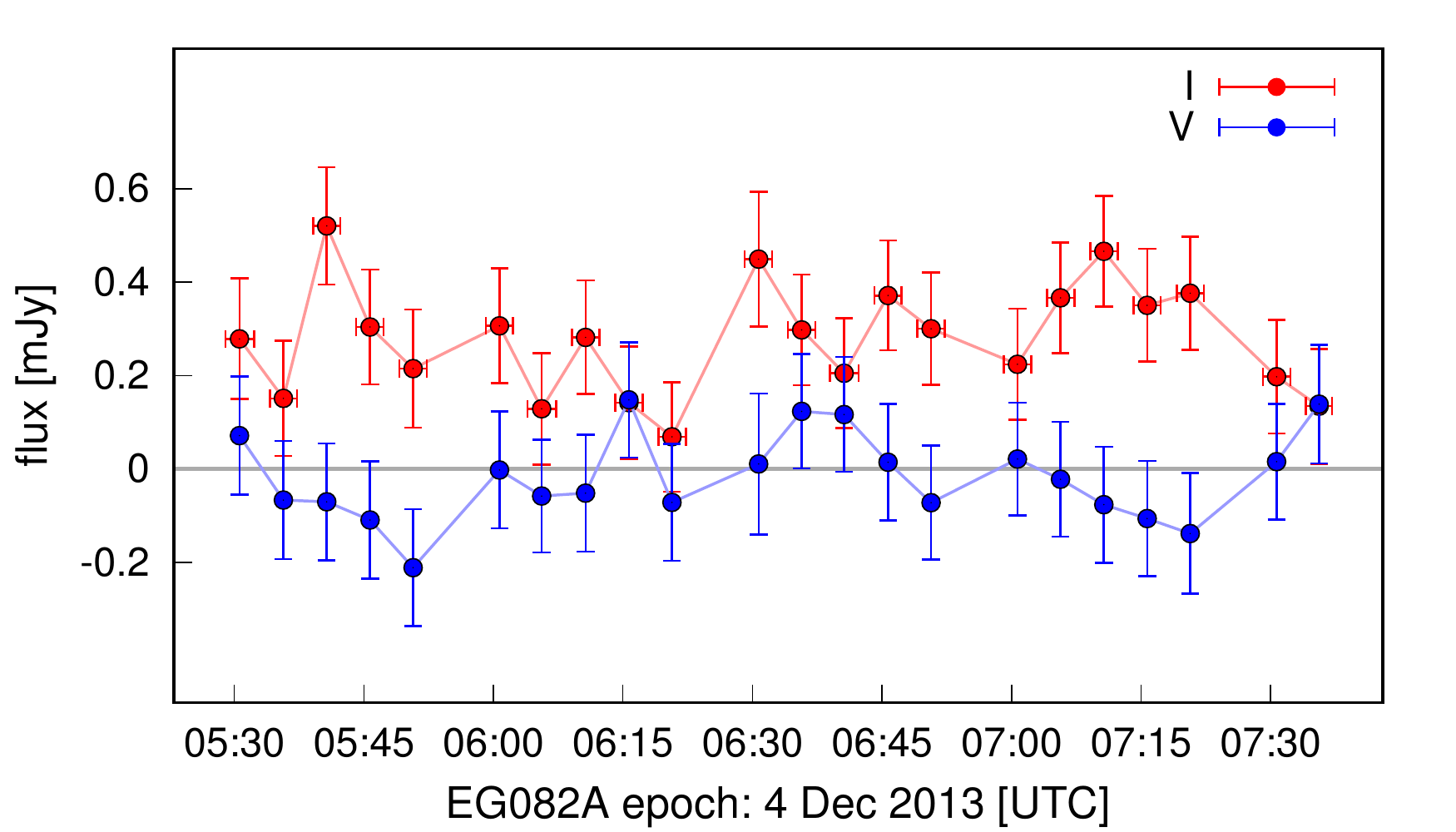}
  \includegraphics[angle=0,width=8.3cm]{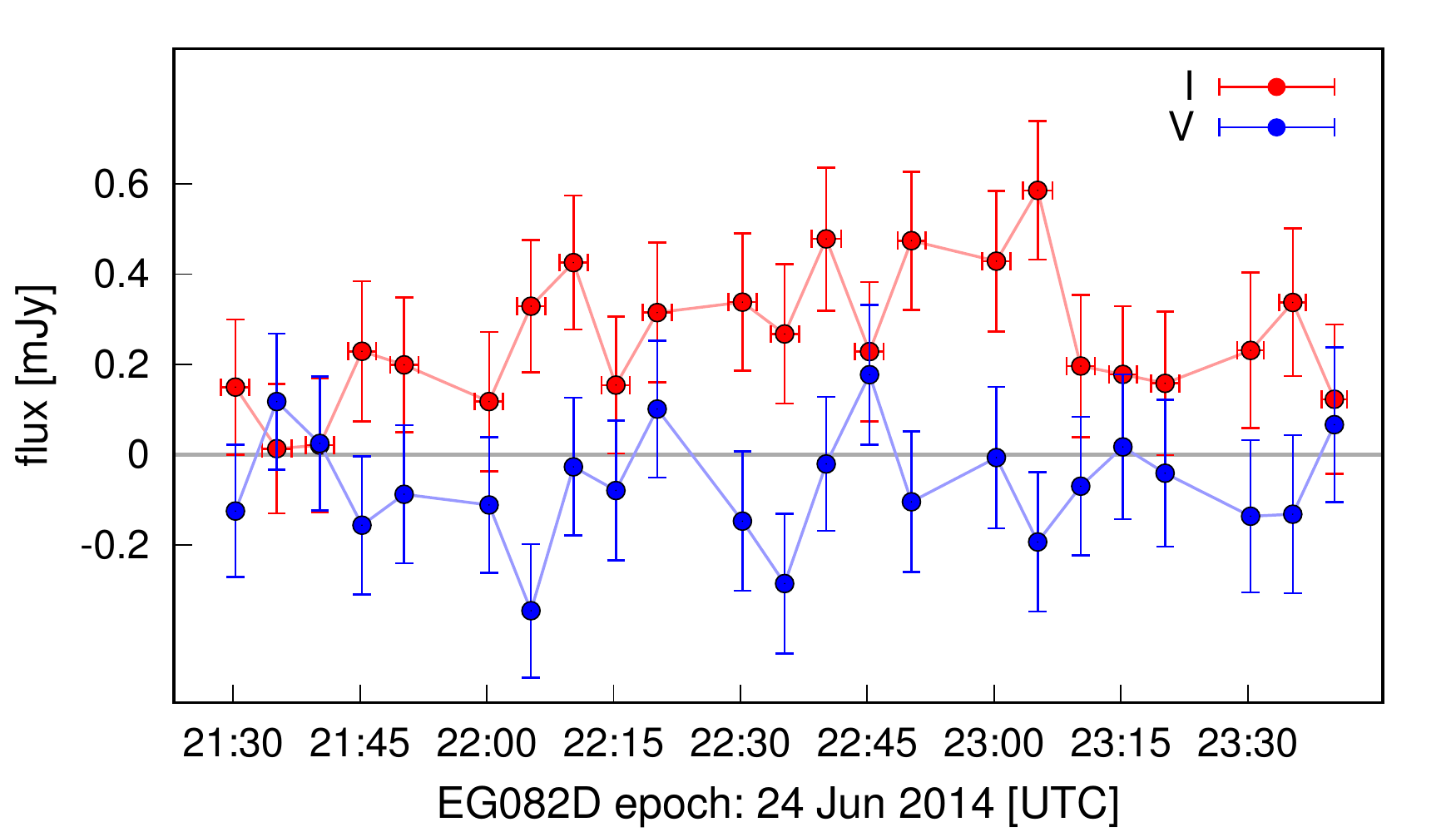}
} 
\caption{
Observed \tvlm{} light curves based on observations from RISARD project. The
total intensity (Stokes $I$, red colour) and circularly polarized (Stokes $V$,
green colour) radio flux at 4.99\,GHz is presented. Flares of right circularly
polarized emission (positive $V$ values) and left circularly polarized emission
(negative $V$ values) are detected. The error bars represent 1$\sigma$ error for
flux and the length of individual integrations during the phase-referencing
observations for time (identical for $I$ and $V$ measurements). Right circular
polarization is is represented by positive $V$ values, and left circular
polarization is represented by negative $V$ values. Lines connecting
the measurement points are shown merely to guide the viewer's eye.
}
\label{fig:fig8}
\end{figure*}

\begin{figure}
\centerline{
\includegraphics[width=0.48\textwidth]{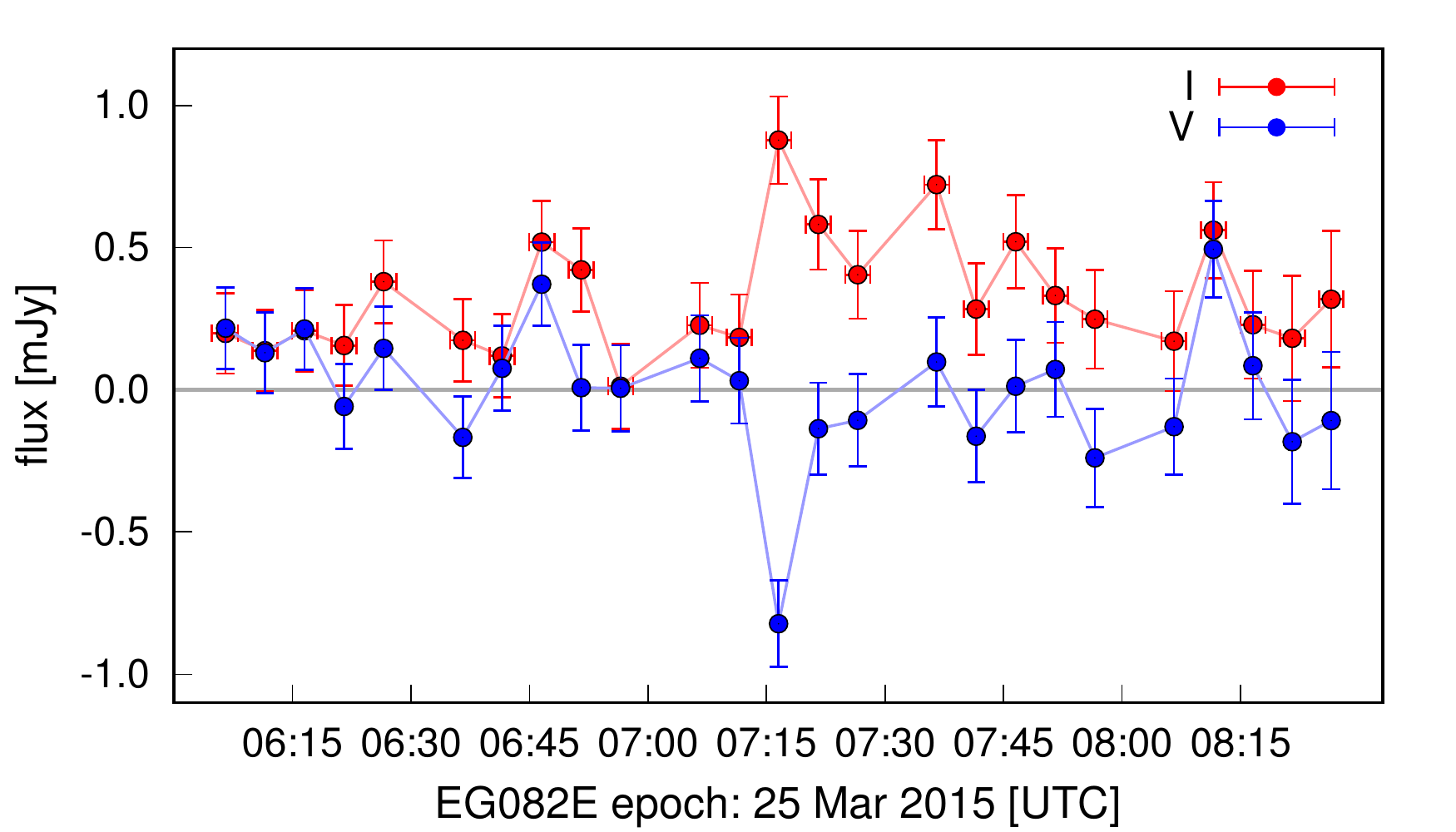}
}
\caption{\tvlm{} light curves continued, see Fig.~\ref{fig:fig8}.}
\label{fig:fig9}
\end{figure}

Previous observations showed that the observed \tvlm{} radio emission consists
of two components, the persistent emission and bursts of highly circularly
polarized radiation \citep[e.g.,][]{2007ApJ...663L..25H}. In order to trace the
highly variable component, we calculated averaged values of Stokes parameters
$I$ and $V$ for each individual scan over the \tvlm{} source during phase-referencing
observations (3.5 min integrations have been used). Such averaging was chosen to
achieve reasonable sensitivity for both parameters. All reconstructed light
curves are presented in Figs.~\ref{fig:fig8} and~\ref{fig:fig9}. We assumed that a flare is detected
when the absolute value of Stokes parameter $V$ is above $2\sigma$ limit. In
addition to the quiescent emission, we detected four short-duration events
($\Delta t\sim3$~min), three left-circular polarization flares at epochs EG065D,
EG065E \& EG082E, and one right-circular polarization flare at epoch EG082E.
Also we observed one broader increase of the flux in the left-circular
polarization spanning $\sim$10~min (EG053b). Our observations make it possible
to detect flares to within $3\sigma$ sensitivity of 0.6\,mJy during an averaged
3.5~min integration. The peak flux density of circularly polarized burst range
from 0.5 to 2\,mJy, with overall fractions of circular polarization
$\sim75\%$--\,$100\%$. 

It is accepted that the quiescent radio emission appears due to the
gyrosynchrotron radiation \citep[e.g.,][]{2006ApJ...647.1349O}. The short
duration of radio flares and high circular polarization suggest coherent process
and electron cyclotron maser was proposed as its likely source
\citep[e.g.,][]{2008ApJ...684..644H}. The electron cyclotron maser radiation is
emitted at the electron cyclotron frequency $\nu_c\simeq 2.8\times 10^{6}B$\,Hz,
where $B$ is the strength of magnetic field in the radio emission region. We
conducted observations at 4.99\,GHz and this frequency infers the small-scale
magnetic field strength of $B\simeq1.8$\,kG. The observed flares are mostly 
left-circularly polarized, in a good agreement with other published 
observations at 4.9\,GHz \citep{2007ApJ...663L..25H}. This supports the model
proposed by these authors to explain periodic flares with period
$P=1.96$\,hr that reflects the rotation period of \tvlm{}. In this model, 
\tvlm{} generates broadband, coherent radio emission in the presence of 
kG magnetic field in a stable, large-scale configuration. 

%
\section{Conclusions}
%
In this work, we present new radio observations of the M9 ultra-cool dwarf
\tvlm{}, using the EVN at 4.99\,GHz. These observations were conducted
in the framework of our RISARD survey \citep{2013arXiv1309.4639G}. \tvlm{} has
been detected at all seven scheduled epochs between March 2011 and March 2015.
It proves an excellent performance and sensitivity of the EVN. 

Combining earlier astrometric data from \citet{Forbrich2009} and
\citet{Forbrich2013} with our
measurements, we updated the astrometric model and the annular parallax
$\pi=93.27^{+0.18}_{-0.17}$\,mas. Unfortunately, the measurements sampling is
sparse, and EVN data are systematically less accurate than VLBA data gathered by
\citet{Forbrich2013}. Therefore we could not detect any clear pattern of the
residuals to the free-falling motion of the target. The irregular residuals pattern make
it possible to rule out putative companions more massive than Saturn in
$\gtrapprox7$~yr orbit or Jupiter in $\gtrapprox1$~yr orbit. The astrometric
positions exhibit a significant error floor $\sim 0.43$~mas, which is comparable
with the astrometric model residuals. This may suggest that the target is either
a single object, either a putative, yet unresolved Jupiter-mass range companion
is present in a short-period orbit (up to one year) contributing to the
apparent, residual noise. Revealing the presence of such a companion would need 
however much dense sampling of the astrometric positions and better accuracy than 
at present can be reached with the EVN. 

Yet the accuracy of derived astrometric positions of \tvlm{} is
comparable with the expected {\sc GAIA} mission outcome. Our results could be a
good reference as an independent observational experiment and will make it
possible to better determine the proper motion and a potential geometric
curvature of the target motion. 

The gathered observational data show a variability of the radio flux on long--
and short--time scales, consistent with published data in earlier papers
\citep[e.g.,][]{2008ApJ...673.1080B}. We detected four highly circularly polarized radio
flares. The short durations of the flares and degrees of circular polarization
are indicative of a coherent emission, which most likely emerges due to the
electron cyclotron maser mechanism. In this context, the inferred local magnetic
field is about 1.8\,kG, similar to values found for other low-mass and fully
convective stars \citep[e.g.,][]{2010MNRAS.407.2269M}. 

%
\section*{Acknowledgments}
%

We thank to the anonymous reviewer for his/her careful reading of the manuscript
and for critical and informative comments which greatly improved and corrected
this work. We are grateful to Polish National Science Centre for their support
(grant no. 2011/01/D/ST9/00735). The EVN is a joint facility of European,
Chinese, South African, and other radio astronomy institutes funded by their
national research councils. K.G. gratefully acknowledges the Poznan
Supercomputing and Networking Centre (PCSS, Poland) for computing grant No. 195.
This research has made use of the SIMBAD database operated at CDS, Strasbourg,
France.

\bibliographystyle{mn2e}  
\bibliography{ms} 
\bsp

\label{lastpage}

\end{document}